\documentclass[acmsmall]{acmart}

\usepackage{amsfonts}

\usepackage{amssymb,mathrsfs,amsmath}
\usepackage{algorithmic}
\usepackage{algorithm}
\usepackage{array}
\usepackage{textcomp}
\usepackage{stfloats}
\usepackage{url}
\usepackage{verbatim}
\usepackage{graphicx}
\usepackage{multirow} 
\usepackage{subcaption}
\usepackage{enumitem}

\usepackage{booktabs}
\usepackage{graphicx}
\usepackage{bbding}
\usepackage{color}
\usepackage{pifont}
\usepackage{tikz}
\usepackage{etoolbox}
\usepackage{tcolorbox}
\usepackage{threeparttable}

\usepackage{listings}
\usepackage{xcolor}

\newtcolorbox{knowledgebox}{
  colback=blue!2!white, 
  colframe=blue!50!black, 
  fonttitle=\bfseries,
  fontupper=\small,
  boxsep=2pt, 
  left=4pt, 
  right=4pt, 
  top=2pt, 
  bottom=2pt 
}

\newcommand{\bigding}[1]{\scalebox{1.2}{\ding{#1}}}

\AtBeginDocument{%
  }

\begin{document}

\setcopyright{acmlicensed}
\acmJournal{TODAES}
\acmYear{2025} \acmVolume{1} \acmNumber{1} \acmArticle{1} \acmMonth{1}\acmDOI{xx.xxxx/xxxxxxx}
\title{Dataset Construction for Training LLM to Learn Analog Circuit Knowledge}

\author{Zihao Chen}
\authornote{These authors contributed equally to this paper.}
\affiliation{%
 \institution{Fudan University}
  \city{Shanghai}
  \country{China}}

\author{Ji Zhuang}
\authornotemark[1]
\affiliation{%
 \institution{Fudan University}
  \city{Shanghai}
  \country{China}}

\author{Jinyi Shen}
\affiliation{%
 \institution{Fudan University}
  \city{Shanghai}
  \country{China}}

\author{Xiaoyue Ke}
\affiliation{%
 \institution{Fudan University}
  \city{Shanghai}
  \country{China}}

\author{Xinyi Yang}
\affiliation{%
 \institution{Fudan University}
  \city{Shanghai}
  \country{China}}

\author{Mingjie Zhou}
\affiliation{%
 \institution{Fudan University}
  \city{Shanghai}
  \country{China}}

\author{Zhuoyao Du}
\affiliation{%
 \institution{Fudan University}
  \city{Shanghai}
  \country{China}}

\author{Xu Yan}
\affiliation{%
 \institution{Fudan University}
  \city{Shanghai}
  \country{China}}

\author{Zhouyang Wu}
\affiliation{%
 \institution{Fudan University}
  \city{Shanghai}
  \country{China}}

\author{Zhenyu Xu}
\affiliation{%
 \institution{Fudan University}
  \city{Shanghai}
  \country{China}}

\author{Jiangli Huang}
\affiliation{%
 \institution{Fudan University}
  \city{Shanghai}
  \country{China}}

\author{Li Shang}
\affiliation{%
\institution{Fudan University}
  \city{Shanghai}
  \country{China}}

\author{Xuan Zeng}
\affiliation{%
 \institution{Fudan University}
  \city{Shanghai}
  \country{China}}

\author{Fan Yang}
\authornote{Corresponding author: yangfan@fudan.edu.cn.}
\affiliation{%
 \institution{Fudan University}
  \city{Shanghai}
  \country{China}}


\renewcommand{\shortauthors}{Zihao Chen et al.}

\begin{abstract}
This paper constructs a textual dataset for training large language models (LLMs) to learn analog circuit knowledge and customizes LLM training techniques. 
For dataset construction, 
high-quality textbooks are collected and decomposed into fine-grained learning nodes, 
which are then used to construct structured question-thinking-solution-answer (QTSA) quadruples using a multi-agent framework to capture both final answers and thought processes. 
The resulting dataset consists of 7.26M tokens of unlabeled data for continual pre-training (CPT) and 112.65M tokens of labeled data for supervised fine-tuning (SFT). 
We customize the training techniques including initial model selection, training paradigms, regularization techniques, and practical implementation references. 
Instruct models are identified as suitable training initialization points, 
an SFT-centric training paradigm is established (finding that CPT provides marginal benefits compared with SFT due to imbalanced data distribution), 
and SFT with KL divergence regularization can achieve a 2.71 percentage-point improvement over SFT alone. 
A practical training implementation method is provided for resource-constrained scenarios. 
Experiments demonstrate that the dataset and training techniques enhance LLMs' analog circuit knowledge.
The trained 32B instruct model achieves 84.59\% accuracy on the AMSBench-TQA benchmark, 
showing a 15.67 percentage-point improvement over the initial model. 
The trained model also shows capability in the operational amplifier design task based on the Atelier framework.
\end{abstract}




\maketitle

\section{INTRODUCTION} \label{section:intro}
Analog circuit design is a challenging field, primarily due to the knowledge-intensive nature of the domain.
This knowledge spans multiple levels: from fundamental device physics and circuit laws, through the analysis and design of basic building blocks such as operational amplifiers (op-amps) and filters, to the architecture of large-scale systems such as phase-locked loops (PLLs) and analog-to-digital converters (ADCs).
Mastering this body of knowledge typically requires years of study and hands-on practice, making analog circuit design a knowledge-intensive field.

Recently, large language models (LLMs), such as the GPT series \cite{gpt3, gpt4, gpt4o}, DeepSeek series \cite{deepseek-v3, deepseek-r1}, and Qwen series \cite{qwen, qwen2, qwen3}, have demonstrated remarkable capabilities in acquiring and applying vast amounts of human knowledge from textual data.
Given that analog circuit design is a knowledge-intensive task, LLMs emerge as promising tools to assist in this domain.
Many works have begun exploring LLM-assisted analog circuit design.
For example, training LLMs on curated datasets has proven effective for designing particular circuits such as op-amps \cite{artisan, autocircuitrl} and passive power converters \cite{ledro}, while multi-agent frameworks equipped with retrieval-augmented generation (RAG) \cite{rag} have empowered general-purpose LLMs in various scenarios including topology design and transistor sizing \cite{analogcoder, atelier, ampagent, analogxpert, anaflow}.

Building upon the encouraging results of these task-specific approaches, we are motivated to construct a textual dataset covering analog circuit knowledge for training LLMs to learn.

However, \textbf{accessible textual training data for analog circuit knowledge is scarce}.
Unlike the software community, where open-source platforms like GitHub and Stack Overflow provide abundant data for LLM training, the analog circuit domain has fewer publicly accessible textual resources.
Existing efforts in dataset construction, such as the AMSNet series \cite{amsnet,amsnet2,amsnetkg} focusing on visual-to-text transformation, and works like \cite{cktgen, cktgnn} focusing on design variable-performance mappings, have made valuable contributions to the field. However, there remains a need for comprehensive knowledge datasets specifically tailored for LLM training.

Moreover, \textbf{effectively training LLMs in this domain requires careful customization of training techniques.}
Training strategies vary substantially across domains. For example, code generation typically benefits from continual pre-training (CPT) \cite{cpt} on massive code repositories \cite{codellama, codebert}, while research in mathematics \cite{math1, math2} and chemistry \cite{chemical} suggests that supervised fine-tuning (SFT) \cite{sft} plays a more critical role compared with CPT.
Given the limited prior experience in this domain, systematically exploring training approaches, including initial model selection, training stage emphasis, algorithm customization, and efficient implementation under resource constraints, requires extensive efforts and computational resources.

In this paper, we construct a domain dataset and customize training techniques for LLMs learning analog circuit knowledge.
Our work consists of two main parts: \textbf{dataset construction} and \textbf{training techniques customization}.

\textbf{For dataset construction:} 
\bigding{172} We first collect high-quality textbooks as the raw corpus. Following the learning roadmap of this field, we gather 20 textbooks spanning circuit theory, analog circuit fundamentals, analog integrated circuit design, and advanced topics.
\bigding{173} However, raw textbook content is unstructured and unlabeled, which is usually insufficient for effective LLM training. To address this, a learning node-based data construction method is proposed to decompose the corpus into fine-grained learning nodes and employ a multi-agent framework to extract structured knowledge in a question-thinking-solution-answer (QTSA) quadruple format, capturing both final answers and underlying thought processes.
\bigding{174} The resulting dataset consists of 7.26M tokens of unlabeled data for CPT and 112.65M tokens of labeled data for SFT.

\textbf{For training techniques customization:} we investigate training strategies through extensive ablation studies and theoretical analysis, addressing four critical aspects: initial model selection, training paradigm emphasis, algorithm customization, and practical implementation under resource constraints.
\bigding{172} For model selection, we compare base, reasoning, and instruct models, identifying instruct models as the most suitable initialization points through both theoretical analysis and empirical validation, achieving over 15 percentage-point performance improvements over the initial model on the benchmark.
\bigding{173} For training paradigm, we investigate the effectiveness of CPT versus SFT-centric approaches. Given the imbalanced data distribution in our dataset, we establish an SFT-centric training paradigm, finding that CPT provides marginal benefits when followed by SFT.
\bigding{174} For algorithm customization, we explore incorporating KL divergence regularization into SFT to improve knowledge learning outcomes. Experiments show that this approach achieves a 2.71 percentage-point improvement over traditional SFT.
\bigding{175} For practical implementation, we provide strategies for resource-constrained scenarios using an asymmetric dual-model deployment technique.
Through the above investigation, we document our preliminary findings and lessons learned regarding these training aspects, hoping they can provide useful references for researchers working on similar challenges. 

In the experiments, a Qwen2.5-32B-Instruct model trained on the proposed dataset achieves 84.59\% accuracy on the AMSBench-TQA benchmark \cite{amsbench}, showing a 15.67 percentage-point improvement over the initial model, which validates the effectiveness of the constructed dataset and training techniques.
The trained model also demonstrates capability in downstream applications by completing the operational amplifier design task in \cite{atelier}.

These contributions are summarized as follows:
\begin{itemize}[leftmargin = 10pt]
\item 
A textual dataset is constructed for training LLMs to learn analog circuit knowledge. 
The dataset is built from high-quality textbooks through a learning node-based data construction method that extracts structured QTSA quadruples using a multi-agent framework, resulting in 7.26M tokens of CPT data and 112.65M tokens of SFT data.
\footnote{ To comply with the copyright terms of the source materials, 
the raw textbook files and the CPT corpus are not distributed.
Researchers may obtain a copy of the synthesized SFT dataset by contacting the corresponding author via email and signing a non-disclosure agreement, under the strict condition of academic use only and no redistribution.}
\item 
Training techniques are customized through ablation studies and theoretical analysis, 
considering four aspects: model selection, training stage emphasis, algorithm customization, and practical implementation under resource constraints. 
The findings and lessons learned are documented in this paper to provide references for researchers in our community facing similar challenges.

\item 
The dataset and training techniques are validated through training experiments, demonstrating that they can enhance LLMs' analog circuit knowledge, achieving 84.59\% accuracy on AMSBench-TQA with a 15.67 percentage-point improvement over the initial model while preserving downstream task execution capabilities in the Atelier op-amp design framework. 

\end{itemize}

The remainder of this paper is organized as follows.  
Section \ref{s2} introduces the background of this work. 
Section \ref{s3} describes the dataset construction process.
Section \ref{s4} presents the investigation of training techniques.
Section \ref{s5} presents the experimental results.
Section \ref{s6} concludes this paper.

\section{BACKGROUND}\label{s2}
\subsection{Knowledge-intensive analog circuit design}\label{s2:analog_background}
Analog circuit design is generally regarded as a demanding discipline in electronic engineering \cite{razavi,martin}.
Designers often need to consider voltage, current, frequency, noise, linearity, and power consumption simultaneously, where these factors are coupled and involve subtle trade-offs.
As a result, practitioners typically need knowledge that spans fundamental physics and mathematics, device-level modeling, circuit-level analysis, and physical implementation.
This knowledge requirement is the main reason why analog circuit design is treated as a knowledge-intensive field.

For example, consider the operational amplifier (op-amp), a representative analog building block. The required knowledge is structured across multiple levels \cite{opampbook,designflow,razavi,martin,nilsson,svoboda2013introduction,alexander2007fundamentals}. 
\textbf{\bigding{172} Fundamental physical laws:}
First, designers rely on physical laws such as Kirchhoff's current and voltage laws and Ohm's law, along with common analysis techniques including nodal analysis, mesh analysis, and Th\'{e}venin/Norton equivalent transformations.
\textbf{\bigding{173} Mathematical grounding:}
Fourier and Laplace transforms are also important, as they support frequency-domain reasoning about circuit behavior.
\textbf{\bigding{174} Transistor device physics and modeling:}
Building on these basics, designers need to understand transistor device physics and associated compact models,
such as the square-law model, small-signal model, and noise model of transistors,
which provide an analytical basis for design decisions.
\textbf{\bigding{175} Single-stage amplifier details:}
With device-level knowledge in hand, designers study amplifier topologies progressively: starting from elementary single-stage configurations (e.g., common-source, common-drain, and common-gate stages) and then moving to structures such as differential pairs, single-ended outputs, folded-cascode, and telescopic-cascode structures.
\textbf{\bigding{176} Multi-stage amplifier design:}
Moving to multi-stage amplifier design introduces additional challenges in frequency response and stability analysis, including pole-zero locations, gain-bandwidth trade-offs, and phase margin considerations. 
\textbf{\bigding{177} Frequency compensation techniques.}
To maintain stable operation, designers often use compensation strategies \cite{opampbook,review} such as Miller compensation, feedforward compensation, and nested feedback compensation, each with its own applicability constraints and design implications.
\textbf{\bigding{178} Physical implementation:}
Finally, translating the schematic into a physical layout adds another layer of considerations, including matching strategies, parasitic-aware design, and placement and routing techniques.

This example illustrates that analog circuit design involves extensive knowledge,
which motivates us to construct a domain-specific knowledge dataset and train an LLM to acquire part of it.

\subsection{Related works of LLM for analog circuit design} \label{s2:related_work_background}
The application of LLMs to analog circuit design automation has emerged. 
LLM applications in topology and parameter design, as the primary decision-making stages in design flow,
are presented in this section. 
Meanwhile, LLMs' roles in other implementation and verification tasks, including executive steps like robust netlist generation, layout optimization, and testbench generation, are also presented. 
Finally, the progress of specialized datasets and benchmarks is summarized, 
revealing the data-scarcity in textual domain-knowledge learning that this work addresses.

\subsubsection{Design task execution.}
Early LLM-assisted analog design efforts focus on netlist generation.
Works such as \cite{analogcoder,analogcoderpro,spicepilot} formulate netlist construction as \texttt{Python} code synthesis with \texttt{PySpice}.
By combining multi-agent coordination and in-context learning, 
these methods report improved netlist generation accuracy and simulation readiness.

Other works focus on topology and parameter design. 
These works guide the LLM to conduct pre-defined topology and parameter design processes with existing design recipes to improve the interpretability and design quality,
either through dataset construction \cite{artisan,lamagic} 
or agentic retrieval from knowledge bases \cite{ampagent,atelier,analogxpert}.
LLM-assisted parameter optimization is recently extensively studied.
References \cite{adollm,llmuso,ledro,easysize} enhance existing black-box methods with LLM guidance, while references \cite{eesizer,liu2025llm,anaflow,autosizer,whiteop} build sizing workflows driven by the reasoning of agents.
Recent works \cite{autocircuitrl,eva} explore combining reinforcement learning with LLMs to explore new topologies.

LLMs are also applied to more downstream tasks. 
For verification, reference \cite{analogtester} automatically construct executable testbenches for given designs. 
For physical implementation, reference \cite{layoutcopilot} builds a multi-agent framework to operate layout tools and supports placement optimization.

\subsubsection{Datasets and benchmarks.}
Applying LLMs to analog circuit design requires high-quality training data, which remains scarce. 
Prior works have produced several datasets and benchmarks, covering various directions: schematic understanding, layout generation, and numerical performance mapping from design variables.

Several efforts focus on visual-to-textual modality transformation.
In the AMSnet dataset series \cite{amsnet,amsnetkg},
schematic images are paired with machine-readable netlists and functional annotations. 
AMSnet 2.0 \cite{amsnet2} further introduces an image segmentation method to recover component positions, improving the schematic parsing robustness and scaling the number of visual-to-textual data samples.
On the other hand, the AMSbench \cite{amsbench} benchmark not only contains subsets to test the multimodal LLMs' (MLLMs) capabilities of translating visual circuit information into text, but also includes a text-only question answering subset (AMSbench-TQA) for testing the analog design knowledge without visual input and thereby providing a testing ground for this work.
The CircuitSense benchmark \cite{circuitsense} evaluates more complex symbolic reasoning capabilities of MLLMs in circuit understanding from schematic images, revealing a critical gap between visual schematic parsing and further symbolic reasoning.

For layout generation,
OSIRIS \cite{osiris} is one of the few available resources that directly target LLM-based layout generation from schematic inputs.
It provides a pipeline that generates clean, verified circuit variations with layout, parasitic-extracted performance metrics, and design parameters.

Other works \cite{cktgnn,cktgen} focus on learning topology-to-performance mapping for other types of neural networks, not tailored for LLMs.
The OCB dataset \cite{cktgnn} offers op-amp instances with circuit topologies and performance labels. It is designed for graph neural networks to learn topology-to-performance mapping or prediction \cite{cktgen}. No natural language annotations are included.

On the other hand, the dataset proposed in this paper does not target a specific design task or step. 
Instead, it is designed to help LLMs learn theoretical knowledge of analog circuits. 
In this knowledge-intensive domain, such understanding can help downstream tasks. 
Besides, this work also investigates the training techniques for LLMs learning analog circuit knowledge.
Therefore, this work is different from existing resources, 
and together they enrich the data spectrum for LLM-assisted analog circuit design.

\subsection{Continual Pre-training of LLMs}
LLMs that have been trained on generic web-scale corpus can be further adapted by a self-supervised phase on unlabeled text.
This procedure, called continual pre-training (CPT) \cite{cpt}, specializes the model while retaining its general abilities.
When the continued corpus is domain-specific, CPT can be referred to as domain-adaptive pre-training (DAPT) \cite{dont_stop_cpt}.

Let $\mathcal{D}_{\text{CPT}}=\{x^{(i)}\}_{i=1}^{N_{\text{CPT}}}$ be the continued corpus, 
where each sequence $x^{(i)}\!=\!(x^{(i)}_1,\dots,x^{(i)}_{N_i})$ contains $N_i$ tokens produced by the same tokenizer as the base model, thus ensuring vocabulary compatibility.  
For a decoder-only architecture such as Qwen series \cite{qwen, qwen2, qwen3},
we minimize the auto-regressive negative log-likelihood loss function:
\begin{equation}
\mathcal{L}_{\text{CPT}}(\theta)
= -\sum_{i=1}^{N_{\text{CPT}}}\sum_{k=1}^{N_i}
      \log p_\theta (x^{(i)}_k \mid x^{(i)}_{<k}),
\label{eq:cpt}
\end{equation}
where $\theta$ denotes the trainable parameters in this model, $x^{(i)}_{<k}$ is the left context $(x^{(i)}_1,\dots,x^{(i)}_{k-1})$, and $P_{\theta}(\cdot|\cdot)$ is the conditional distribution over the vocabulary obtained via a softmax layer.
Minimizing this cross-entropy objective continues the original pre-training regime without injecting task-specific bias.

\subsection{Supervised Fine-tuning of LLMs}
Supervised fine-tuning (SFT) \cite{sft} adapts a pre-trained model to the reasoning style required by specific downstream tasks using labeled examples (nowadays typically formatted as question-answer pairs).

Let $\mathcal{D}_{\text{SFT}}\!=\!\{(x^{(i)},y^{(i)})\}_{i=1}^{N_{\text{SFT}}}$ be the labeled training set, 
where $x^{(i)}\!=\!(x^{(i)}_1,\dots,x^{(i)}_{N_i})$ denotes the $i$-th input sequence (question) with $N_i$ tokens,
and $y^{(i)}\!=\!(y^{(i)}_1,\dots,y^{(i)}_{L_i})$ denotes the corresponding target sequence (answer) with $L_i$ tokens.  
A decoder-only language model with parameters $\theta$ is optimized by the following cross-entropy loss function:
\begin{equation}
  \mathcal{L}_{\mathrm{SFT}}(\theta)=
  -\sum_{i=1}^{N_{\text{SFT}}}\;\sum_{k=1}^{L_i}
        \log p_{\theta}(y^{(i)}_{k}\mid x^{(i)},y^{(i)}_{<k}),
  \label{eq:sft}
\end{equation}
where $\theta$ denotes the trainable parameters in this model, $y^{(i)}_{<k}$ is the left context $(y^{(i)}_1,\dots,y^{(i)}_{k-1})$, and the inner term maximizes the probability of the $k$-th ground truth token conditioned on the complete input and all previously generated target tokens.  
Minimizing Eq. \eqref{eq:sft} aligns the model's output logits with labeled values.

\subsection{Categorization of LLMs from Training Aspects} \label{sec:llm_types}
From a training perspective, LLMs fall into three tiers:

\textbf{(i) Base models.}
LLMs are first trained from random initialization with a purely auto-regressive objective like Eq.~\eqref{eq:cpt} in a corpus with billions to even trillions of tokens \cite{gpt3,llama2,qwen2}.  
This pre-training imparts broad statistical knowledge, yet leaves the models mostly unaligned with human instructions, 
so they cannot reliably follow instructions or produce safe and truthful dialogue \cite{gpt3,gpt4}. 
Therefore, additional SFT and RL from human feedback (RLHF) \cite{rlhf} stages are typically required before final deployment.

\textbf{(ii) Instruct models.}
SFT and RLHF are important and expensive stages in converting a pre-trained base model into a reliable conversational agent, i.e., an instruct model.
This process depends on large, diverse, and meticulously curated instruction and preference data pairs.
For example, the Qwen2.5 series employs millions of SFT and RLHF data, trains for two epochs with sequences up to 32k tokens, and requires multi-week runs on thousands of GPUs \cite{qwen2}.
Although reference \cite{LIMA} shows that roughly ten thousand carefully curated instruction-response pairs can align a model for everyday dialogue, such small-scale supervision still falls short of the performance attained by official models, especially on more demanding tasks.

\textbf{(iii) Reasoning models.}  
Reasoning models are built by adding RL on top of instruct models, typically initialized with SFT and then refined with GRPO \cite{deepseek-r1} or recent DAPO \cite{dapo}. 
This pipeline enhances math and coding performance but requires millions of preference pairs and extensive computational resources.
However, the parameter distribution of reasoning models is usually sparsely redundant and highly sensitive, potentially making them prone to collapse or performance degradation during further fine-tuning \cite{constitutional,dpo}. 
Whether reasoning models are suitable for specific domain adaptation depends on various factors and requires careful investigation. 
\subsection{Typical LLM Domain Customization Pipeline} \label{sec:training_pipline}
Conventional workflow typically begins with CPT to incorporate domain text, 
followed by SFT on instruction data. 
Due to its higher resource demands, RLHF is sometimes excluded in academic research. 
The effectiveness of these stages varies significantly across different domains.

In programming tasks, for example, CPT is typically indispensable.
Modern code models are usually exposed to billions to trillions of raw code tokens \cite{codebert,codellama}.
Such large-scale pre-training yields substantial gains over models that rely on SFT alone, 
suggesting that language-agnostic syntax and cross-language identifier semantics should be absorbed during CPT.

In contrast, mathematics appears far less reliant on CPT. For example, reference \cite{math1} solely trains a 13B model on 5M question-answer pairs, and achieves good performance matching or outperforming code-pretrained models of similar size. Follow-up work \cite{math2} further shows that scaling SFT alone enables even a vanilla 7B model to close most of the remaining gap. 
These findings suggest that general-purpose models already encode latent mathematical structure, which can be effectively unlocked through sufficiently rich SFT.

These mixed outcomes suggest that the training strategy is highly domain-dependent.
The optimal balance between CPT and SFT, as well as the relative importance of each stage, depends on various factors including the domain characteristics, data availability, and the scale and distribution of available training data.

\subsection{KL Divergence Regularization in LLM Training} \label{sec:kl_regularization}
LLMs are fundamentally generative models that output probability distributions over tokens.
When adapting a pre-trained model to new domains, the model's output distribution may shift significantly during fine-tuning.
To control this shift and maintain stability during adaptation, regularization techniques can be employed to constrain the model's output distribution to remain close to the original model.
When considering regularization for LLMs, KL divergence naturally emerges as a principled choice, as it directly measures the difference between probability distributions, i.e., the native output space of LLMs.
The Kullback–Leibler (KL) divergence \cite{kl} quantifies the information loss when approximating one distribution with another:
\begin{equation}
D_{\text{KL}}(p \parallel q) = \sum_{x} p(x) \log \frac{p(x)}{q(x)}.
\label{eq:kl_divergence}
\end{equation}

However, the effectiveness of KL divergence regularization in LLM training is highly context-dependent.
For example, in RL for improving LLMs' reasoning capabilities, while GRPO \cite{deepseek-r1} successfully employs KL divergence regularization term for model training, DAPO \cite{dapo} removes this term and also achieves competitive results.
On the other hand, in SFT for LLMs' domain adaptation, KL divergence regularization is less used and the effectiveness is not well studied for the analog circuit domain.
Consequently, whether KL divergence regularization benefits domain adaptation tasks, particularly depending on the specific characteristics of the domain and training conditions, remains an open empirical question that warrants careful investigation.

\section{Dataset Construction}\label{s3}
This section presents the process of constructing the analog circuit textual knowledge dataset for LLM training,
which consists of two main stages: 
Section \ref{s3a} describes the collection and preprocessing of corpora following the learning roadmap of analog circuits; 
Section \ref{s3b} presents the learning node-based data construction pipeline, 
which consists of two components: 
corpus decomposition into learning nodes (Section \ref{s3b1}) and 
a multi-agent framework for data construction (Section \ref{s3b2}). 
Section \ref{s3c} provides dataset statistics. 

\subsection{Corpus Collection and Preprocessing}\label{s3a}
As discussed in Section \ref{section:intro}, accessible textual training data for analog circuit knowledge is scarce, which has become a challenge for applying LLMs in analog circuit design.
This section first collects textbooks as the raw corpus, following the learning roadmap of analog circuits.

The theoretical learning of analog circuits is usually divided into four stages: circuit theory, analog circuit basis, analog integrated circuits, and advanced circuit topics. 
\textbf{\bigding{172} Circuit theory} encompasses basic circuit laws, network analysis methods, time and frequency domain analysis, and other fundamental knowledge, establishing the mathematical foundation for future learning.
\textbf{\bigding{173} Analog circuit basis} focuses on device characteristics, basic amplifier circuits, feedback theory, and stability analysis, enabling learners to develop a fundamental understanding of circuit functionality and design techniques.
\textbf{\bigding{174} Analog integrated circuits} delves into the design of core modules such as op-amps, comparators, and reference circuits, as well as the implementation of circuits under CMOS technology.
\textbf{\bigding{175} Advanced circuit topics} covers specialized circuits from op-amps to PLLs.

\setlength\tabcolsep{3pt}
\renewcommand{\arraystretch}{1.1}
\begin{table}[!h]
    \centering
    \caption{Details of Collected Analog Circuit Corpus}
    \label{table:corpus}
    \begin{tabular}{c|c|c}
        \hline
        \multicolumn{3}{c}{\textbf{Part 1: Books Categorized by Learning Stage}} \\ \hline
        \textbf{Learning stage} & \textbf{Circuit scope} & \textbf{Book number} \\ \hline
        Circuit theory & Basic passive components and circuits & 1 \\ \hline
        Analog circuit basis & Basic amplifiers and active filters & 3 \\ \hline
        Analog integrated circuits & Various circuits with various scales & 5 \\ \hline
        Advanced circuit topics & Focused circuits for specific applications & 11 \\ \hline
    \end{tabular}
    \vspace{1.5em}
    
    \begin{tabular}{c|c|c|c}
        \hline
        \multicolumn{4}{c}{\textbf{Part 2: Books Categorized by Circuit Types with Overlap }} \\ \hline
        \textbf{\begin{tabular}[c]{@{}c@{}}Circuit type\end{tabular}} & \textbf{\begin{tabular}[c]{@{}c@{}}Subtype number \end{tabular}} & \textbf{\begin{tabular}[c]{@{}c@{}} Subtype example(s)\end{tabular}} & \textbf{\begin{tabular}[c]{@{}c@{}}Book number \end{tabular}} \\ \hline
        Basic circuits & - & - & 4 \\ \hline
        Amplifiers & 6 & Op-amp, power amplifier & 9 \\ \hline
        Arithmetic circuits & 4 & Multiplier, divider & 5 \\ \hline
        Filters & 2 & Active filter & 6 \\ \hline
        Power management circuits & 2 & DC-DC, AC-DC & 2 \\ \hline
        Analog-to-Digital converters & 8 & SAR ADC, $\Sigma\text{-}\Delta$ ADC & 6 \\ \hline
        Digital-to-Analog converters & 2 & $\Delta\text{-}\Sigma$ DAC & 6 \\ \hline
        Reference circuits & 3 & Voltage reference & 5 \\ \hline
        Regulators & 5 & Linear regulator & 5 \\ \hline
        Transceivers & 4 & Transmitter, modulator & 5 \\ \hline
        Oscillators & 3 & VCO, LC oscillator & 5 \\ \hline
        PLLs & 1 & Charge-pump PLL & 6 \\ \hline
    \end{tabular}
\end{table}

As shown in Table \ref{table:corpus} Part 1, for the first three stages, we select several textbooks with complete content systems, totaling 9 volumes. 
For the advanced circuit design stage, we select 11 reference books to cover core design techniques in the major application areas.
Therefore, we construct a corpus containing 20 core textbooks legitimately for only academic use, which together span at least 12 major circuit types with overlaps, as detailed in Table \ref{table:corpus} Part 2, from simple to complex.

At the implementation level, the commercial parser \texttt{Mathpix} is employed to convert the original PDF files into a clean Markdown-formatted corpus, including text recognition, structured processing of mathematical formulas and tables.
Subsequently, a combination of automated scripts and manual adjustments is used to complete the desensitization and denoising of the corpus, finally obtaining an unannotated corpus containing 7.26M tokens.

It should be noted that this method does not incorporate visual modalities in the corpus.
On the one hand, this strategy aligns with the development of LLMs in other specialized domains \cite{dont_stop_cpt,scibert}. 
For example, early-stage LLMs in medicine \cite{biomedical}, law \cite{legal}, and mathematics \cite{llemma} also relied primarily on domain-specific textual knowledge injection and gained performance improvements on textual benchmark tasks, without incorporating visual modalities.
On the other hand, in the analog circuit domain, 
circuit images contain numerous fine-grained and specialized visual details. 
The processing techniques for schematic images remain under development \cite{amsnet,amsnet2,amsnetkg,circuitsense,ocr3,sina,doceda}.
Moreover, visual information in analog circuit textbooks extends far beyond schematics 
to include photographs, function waveforms, device structure diagrams, and more.
A stable, general-purpose image-to-text mapping for the full variety of analog circuit visuals remains an open challenge. Developing such methods is beyond the scope of this paper but is an important direction for future research.
Experimental results will demonstrate the effectiveness of this data construction approach.

\subsection{Learning Node-based Data Construction}\label{s3b}
Textbook corpus is unstructured and unlabeled, which is insufficient for LLM training. 
A learning node-based analog circuit data construction method is proposed, which consists of two steps: 
decomposing the corpus into fine-grained learning nodes (Section \ref{s3b1}) 
and employing a multi-agent framework to construct labelled data pairs (Section \ref{s3b2}).

\subsubsection{Corpus Decomposition into Learning Nodes}\label{s3b1}
The cleaned corpus remains extremely long text files, making it difficult to effectively generate SFT data from it. 
A single book could cover different learning stages, encompassing dozens of circuits with vastly different principles and design considerations. 
Therefore, we need to decompose the corpus into finer-grained units that are both manageable in size and semantically coherent for SFT data construction.

\begin{figure*}[!b]
    \centering
    \includegraphics[width=1\linewidth]{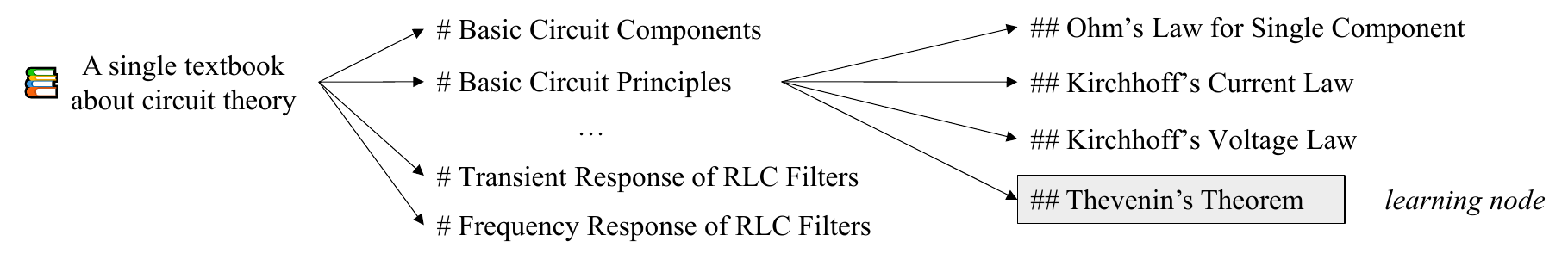}
\caption{An example of the granular decomposition of the cleaned corpus. From left to right, the granularity becomes progressively finer. For each book, the first stage is the section, the second stage is the subsection, i.e. the defined learning nodes.}
\label{fig:decomposition}
\end{figure*}

The collected analog circuit textbooks typically feature a hierarchical structure with two-level subsections. 
As illustrated in Fig. \ref{fig:decomposition}, 
a textbook can be decomposed at multiple granularities: 
from the entire book, to sections (chapters), and to subsections (learning nodes).
For example, a textbook related to circuit theory can usually be divided into many sections, 
from \emph{basic circuit components} to \emph{frequency response of RLC filters}, 
each section being further subdivided into several subsections. 
Based on this observation, we define each subsection as a learning node, 
i.e., the finest granularity unit for SFT data construction.
This is motivated by the fact that subsections represent semantically complete knowledge units: 
they are neither too fine-grained (which would fragment related concepts) 
nor too coarse (which would mix different topics), 
thereby keeping knowledge integrity while being suitable for focused SFT data construction.

The statistics on the collected corpus primarily validate this decomposition strategy. 
Out of all identified subsections, 
we filter out brief, general introductory content and identify 2.69k valid subsections as learning nodes. 
We subsequently analyze the length of each learning node, finding an average of about 2k tokens per node, which is a relatively moderate length suitable for data construction,
neither too short to lack context nor too long to be unwieldy for the subsequent agentic data construction process.

In the implementation, we partition the cleaned corpus according to the extracted chapter and hierarchical information. 
The Markdown corpus preserves the hierarchical chapter structure with multi-level headings (e.g., \# for sections and \#\# for subsections as shown in Fig. \ref{fig:decomposition}).
Automated scripts parse these hierarchical heading structures and split the corpus into learning nodes.

\subsubsection{Multi-agent Framework for Data Construction}\label{s3b2}
\begin{figure*}[!h]
    \centering
    \includegraphics[width=1\linewidth]{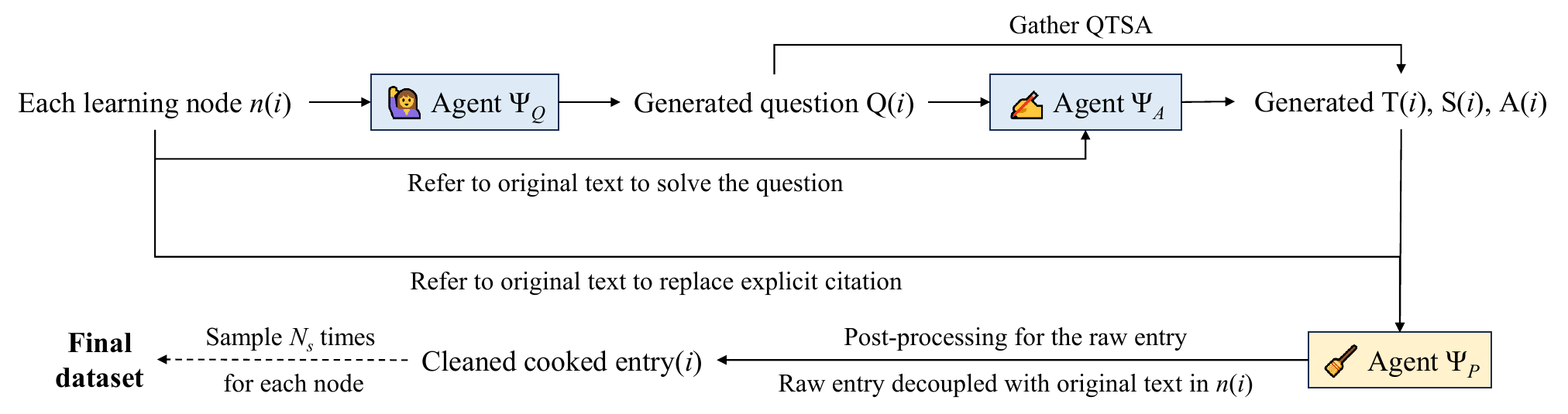}
\caption{The multi-agent framework for data construction. 
The upper half is the data generation stage, while the lower half is the post-processing stage.}
\label{fig:data_gen}
\end{figure*}

Structured data pairs are constructed from each learning node using LLM agents. 
Note that knowledge within a learning node may involve not only memorization and comprehension but also reasoning or problem-solving.
Therefore, a multi-agent framework consisting of three reasoning agents ($\Psi_Q$, $\Psi_A$, and $\Psi_P$) is employed to construct SFT data pairs.
As illustrated in Fig. \ref{fig:data_gen}, the framework operates in two main stages: 
a data generation stage (upper half) where questions are generated and answers are produced, 
and a post-processing stage (lower half) where the generated data is cleaned and refined.

To generate data pairs containing both the final answers and the thinking processes, 
a \emph{question-thinking-solution-answer} (QTSA) quadruple is adopted as the data format.
For each learning node $n(i)$, the data entry is defined as:
\begin{equation}
\text{entry}(i) = [\text{Q}(i), \text{T}(i), \text{S}(i), \text{A}(i)],
\end{equation}
where 
$\text{Q}(i)$ represents the generated question, 
$\text{T}(i)$ is the thinking process, which includes informal analysis of the question as well as the entire process of formulating the final solution and answer, 
$\text{S}(i)$ is the refined and relatively structured solving process, 
and $\text{A}(i)$ is the final answer.
For training, each data entry $\text{entry}(i)$ is converted into an $(x^{(i)}, y^{(i)})$ pair following the standard SFT format (as defined in Section \ref{sec:training_pipline}), where the input $x^{(i)} = \text{Q}(i)$ and the output $y^{(i)}$ consists of $\text{T}(i)$, $\text{S}(i)$, and $\text{A}(i)$ concatenated together, with XML separators (such as {\small \textless}think{\small \textgreater}...{\small \textless}/think{\small \textgreater} and {\small \textless}answer{\small \textgreater})...{\small \textless}/answer{\small \textgreater} used to delimit the different components within $y^{(i)}$.

\begin{figure}[!h]
    \centering
    \includegraphics[width=.77\linewidth]{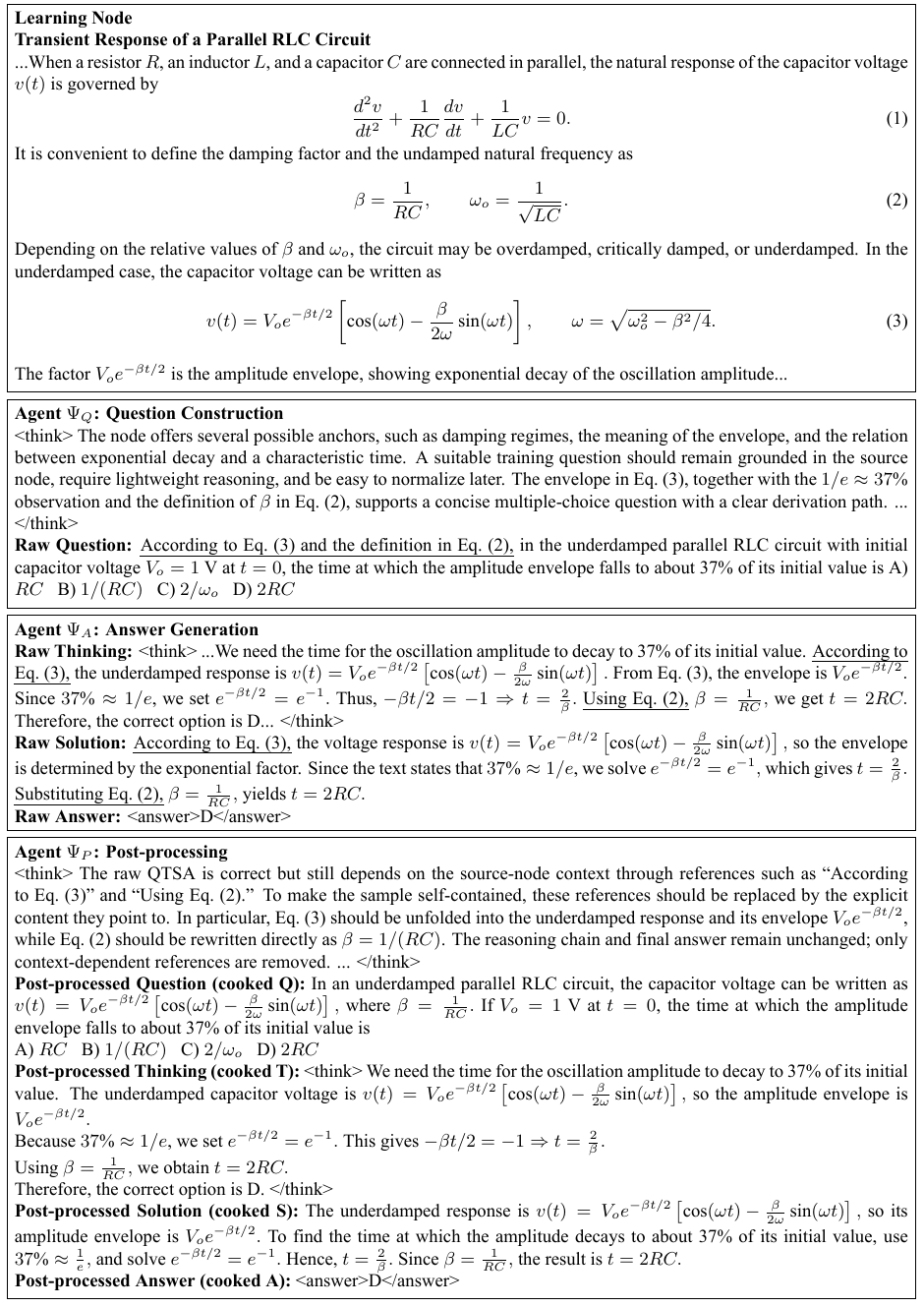}
\caption{An illustrative example of the QTSA data construction process. Some content is abbreviated for brevity.}
\label{fig:qtsa}
\vspace{-1em}
\end{figure}

As shown in Fig.~\ref{fig:qtsa}, the data generation stage involves two agents working sequentially.
First, for each learning node $n(i)$, agent $\Psi_Q$ generates a question based on the original content of the learning node:
\begin{equation}
Q(i)=\Psi_Q[n(i)].
\end{equation}
The prompt design for $\Psi_Q$ requires it to propose a guided question that reflects the information in $n(i)$ based on the original text content.
Specifically, the question must be grounded in the original text content and should be answerable using the original content with reasoning and calculations.
This design ensures that the generated questions are aligned with the learning node content.
Fig.~\ref{fig:qtsa} shows an example where agent $\Psi_Q$ proposes a question based on the RLC circuit transient response knowledge.

Subsequently, agent $\Psi_A$ produces the answer based on its own knowledge and abilities, the original learning node content $n(i)$ and the generated question $Q(i)$:
\begin{equation}
\text{entry}(i)=\Psi_A[n(i), Q(i)].
\end{equation}
Specifically, DeepSeek-R1 \cite{deepseek-r1}, a reasoning model that naturally generates both thinking processes and formal answers, is employed as the agent in this data generation framework.
The model automatically generates {\small \textless}think{\small \textgreater}...{\small \textless}/think{\small \textgreater} tags containing the reasoning process, which we extract as $\text{T}(i)$.
The content outside the {\small \textless}think{\small \textgreater} tags constitutes the formal answer, which we extract as $\text{S}(i)$.
Furthermore, the final answer should be wrapped in {\small \textless}answer{\small \textgreater}...{\small \textless}/answer{\small \textgreater} tags, allowing the content within these tags to be extracted as $\text{A}(i)$ from $\text{S}(i)$.
Fig.~\ref{fig:qtsa} shows an example where agent $\Psi_A$ generates the raw thinking, solving process, and final answer.
To ensure comprehensive knowledge coverage in each learning node, we independently sample each learning node $N_S$ times.

The generated data undergoes further postprocessing and cleaning through agent $\Psi_P$,
whose functionality can be seen in Fig. \ref{fig:qtsa}.
First, the agent removes ill-formatted items, 
including those that do not conform to the QTSA structure or contain obvious errors.
Second, it identifies referential expressions in the answers, 
such as references to formulas, figures, or tables from the original corpus, 
and replaces them with explicit content.
For example, when the generated answer contains phrases like "as shown in Eq.~(3)" or "according to the definition in Eq.~(2)" (highlighted with underlines in Fig. \ref{fig:qtsa}), 
the post-processing agent $\Psi_P$ locates the corresponding formula in the original learning node and rewrites the reference as the actual content. 
This decoupling process helps ensure that each sample remains relatively self-contained
even when removed from the original book layout, thereby improving the usability of the dataset.

\subsection{Dataset Statistics}\label{s3c}
Table \ref{table:dataset} summarizes the dataset scale after the automated construction and screening steps. 
The original corpus was decomposed into 2.69k unlabeled CPT data, containing 7.26M tokens. 
Using the multi-agent methods introduced in this section, each learning node was independently sampled repeatedly. 
After filtering out duplicate samples and obviously erroneous samples, 
we obtained 15.31k data entries.
Fig. \ref{fig:qtsa} presents an example of the QTSA data entry.
Compared with other domains, 
the scale of this dataset remains limited. 
Nevertheless, under the experimental settings of this paper, 
the constructed dataset can serve as the input for LLM training.

For data correctness verification, we manually checked 500 randomly sampled instances from the SFT dataset.
Of these, 455 entries (91.00\%) were found to be correct and well-formed. 
The remaining 45 entries, including 2 entries with formatting issues and 43 entries with factual errors or hallucinations.
The performance improvement in subsequent experiments on the AMSBench-TQA test benchmark can serve as evidence of the dataset's effectiveness.

For original textbook content and SFT dataset similarity filtering, 
the cosine similarity between each QTSA sample and its corresponding source learning node text is computed using the sentence embedding model from \cite{reimers2019sentence}. 
Samples with a similarity score exceeding 0.8 are considered as potential contaminated samples, which are further checked manually. 
In total, 236 SFT samples exhibited a similarity score exceeding 0.8 (none of them exceeded 0.9). 
We manually reviewed all 236 entries against the original textbook learning nodes to assess whether they contained near-verbatim reproduction of original textbook content. 
Finally, 140 samples are identified that retained phrasing too close to the original textbook content.
These small amount of samples are removed from the final SFT dataset. 

\setlength\tabcolsep{2pt}
\renewcommand{\arraystretch}{1.2}
\begin{table}[!h]
    \centering
    \caption{Details of the Constructed Dataset}
    \label{table:dataset}
    \begin{tabular}{c|c|c}
        \hline
        \textbf{Training stage} & \textbf{Data entry number} & \textbf{Data token number} \\ \hline
        \textbf{CPT} & 2.69k & 7.26M \\ \hline
        \textbf{SFT} & 15.31k & 112.65M \\ \hline
    \end{tabular}
\end{table}
\section{Training Techniques Customization}\label{s4}
This section presents the training techniques for effectively utilizing the constructed dataset to train an LLM in learning analog circuit knowledge.
Through extensive ablation studies, we investigate training strategies for analog circuit domain adaptation, addressing four aspects that are critical for effective LLM training: 
Section \ref{s4a} investigates the selection of appropriate pre-trained models as initialization points, identifying instruct models as the better choice over base and reasoning models; 
Section \ref{s4b} establishes an SFT-centric training paradigm, demonstrating that CPT provides only marginal benefits; 
Section \ref{s4c} investigates the effectiveness of KL divergence regularization in SFT; 
and Section \ref{s4d} provides practical implementation strategies for resource-constrained scenarios.

\subsection{Model Selection for Training Initialization}\label{s4a}
The selection of an appropriate pre-trained model as the initialization point is crucial for training an LLM in learning analog circuit knowledge. 
As discussed in Section \ref{sec:llm_types}, the current LLM landscape comprises three categories,
each of which presents unique characteristics that impact their suitability for analog circuit knowledge learning.

\textbf{(i) Base models: impractical and unnecessary.}
While base models offer theoretical flexibility for domain customization, we identify that adopting a base model is both impractical and unnecessary in our scenario.
On the one hand, high-quality instruct checkpoints like Qwen2.5-32B-Instruct are already publicly available, leaving expensive retraining from a base model unnecessary.
On the other hand, reproducing training from base models is infeasible, since it requires the proprietary SFT and RLHF datasets used in official post-training, which are not publicly released for leading open-source models \cite{qwen, qwen2, qwen3}.
Therefore, we further investigate whether a reasoning model or an instruct model is more suitable for training an analog circuit LLM.

\textbf{(ii) Reasoning models: superior but fragile.}
While reasoning models excel at complex tasks, they are sensitive to further training on domain-specific data.
For example, a reasoning model such as DeepSeek-R1 \cite{deepseek-r1} is trained through GRPO-like RL methods~\cite{deepseek-r1,dapo}, where the extensive online sampling with reward feedback customizes a model parameter distribution for difficult reasoning tasks \cite{huan2025does,xie2025logic}.
References \cite{kirk2023understanding,policycliff} further demonstrate that RL processes applied to LLMs exhibit brittleness, i.e., 
small parameter changes can lead to large, unpredictable shifts in output. 
This indicates that heavy RL training narrows the model's parameter numerical distribution, reducing its capacity to adapt to further domain-specific fine-tuning \cite{policycliff}.

To assess whether such fragility arises in the context of analog circuit design,
the following ablation study is conducted.
As shown in Table \ref{table:performance_comparison}, fine-tuning QwQ-32B \cite{qwen2}, an open-source reasoning model through SFT, 
results in a performance degradation from 80.78\% to less than 75\% on the benchmark.
This degradation validates our hypothesis and is consistent with prior observations \cite{kirk2023understanding,policycliff,rlvssft} regarding the fragility of RL-optimized model parameter distributions.

\textbf{(iii) Instruct models: aligned and adaptable.}
On the other hand, instruct models provide a more suitable initialization for training LLMs in the analog circuit domain. 
An instruct model, such as Qwen2.5-32B-Instruct \cite{qwen2}, is derived from a base model through extensive SFT followed by relatively lightweight RLHF. 
The RLHF stage of instruct models often achieves instruction following and safety alignment \cite{dpo}, and its impact on the model's parameter distribution is limited \cite{raina2025d} when compared to the GRPO-like RL process to train reasoning models \cite{wu2025takes}. 
This difference in RL intensity accounts for the different behaviors of reasoning and instruct models during further domain-specific fine-tuning.
Recent studies also compare the different roles of SFT and RL. 
References \cite{ding2025rethinking,rajani2025scalpel} indicate that SFT retains the model's ability to learn new tasks, 
while reasoning models become brittle under further domain-specific fine-tuning because heavy RL training overly tightens the model's parameter distribution and exhausts its plasticity \cite{kirk2023understanding,policycliff,rlvssft}.

As shown in Section \ref{s5}, fine-tuning Qwen2.5-32B-Instruct can achieve a performance improvement of more than 15 percentage points, 
validating that analog circuit knowledge can be effectively integrated through the further training of instruct models.

\subsection{SFT-centric Training Paradigm}\label{s4b}
While Section \ref{sec:training_pipline} establishes that training strategies are highly domain-dependent, the optimal configuration for analog circuits remains unexplored and requires investigation.
Analog circuit knowledge is both complex and domain-specific, yet our dataset exhibits a significant imbalance: the corpus contains only 7.26M tokens for CPT versus 112.65M tokens for SFT (approximately 16$\times$ larger).
Prior work suggests that CPT requires substantial data volumes typically exceeding 100M tokens \cite{dont_stop_cpt}, which is well above our CPT corpus size.
Given these characteristics, we hypothesize that:
CPT could provide minimal benefits due to insufficient data scale,
while SFT alone may achieve comparable performance by directly teaching domain knowledge through instruction tuning.

To validate this hypothesis, we conduct a controlled ablation study in Section \ref{s5}, comparing SFT-only versus CPT+SFT pipelines.
Our results confirm that the CPT+SFT pipeline yields only a marginal 0.17 percentage-point improvement over SFT alone (see Table \ref{table:ablation_i}), which could be negligible and likely within the range of random variation.
Therefore, an SFT-centric approach is more efficient for training an analog circuit LLM.

\subsection{SFT with KL Divergence Regularization}\label{s4c}

\subsubsection{The Motivation of Using KL Divergence Regularization}
Given our domain-specific SFT dataset is costly yet limited, 
maximizing the utility of this dataset becomes important. 
Analog circuit knowledge is both complex, 
requiring the model to learn domain-specific knowledge 
while preserving general capabilities of understanding and instruction following.

However, fine-tuning an LLM on new data could cause the model to forget some of its previously learned knowledge~\cite{forget}.
When we inject new analog circuit knowledge into the model (noted as $\Delta K_{\text{inject}}$), 
the model also loses some of its prior knowledge (noted as $\Delta K_{\text{forget}}$). 
For conceptual illustration (not a formal measurement),
the total domain knowledge gain $\Delta K_\text{gain}$ can be roughly represented as:
\begin{equation}
\Delta K_\text{gain} = \Delta K_{\text{inject}} - \Delta K_{\text{forget}}.
\end{equation}

In this work, the analog circuit training dataset is relatively small and difficult. 
Under such conditions, the forgetting term $\Delta K_{\text{forget}}$ can become quite large and harm overall performance. 
Therefore, any technique that reduces forgetting should help increase the total knowledge gain.

A KL divergence regularization term during SFT can help reduce this forgetting. 
At first glance, this may seem counterproductive: 
SFT encourages the model to fit new data with the cross-entropy loss, 
while KL term penalizes deviation from the original model. 
However, the two terms work in a complementary way. 
SFT provides the necessary gradient signal to inject domain knowledge, 
while the KL term acts as a mild regularizer that prevents the model from drifting too far from its pre-trained behavior. 
This combination is widely used in RL~\cite{ppo,dpo,deepseek-r1}, 
where KL penalties can constrain policy updates and maintain training stability~\cite{korbak2022rl,zhang2025design}. 
Recent work \cite{zhu2025anchored} also shows that KL-regularized SFT serves as an anchoring mechanism, mitigating over-memorization and parameter distribution drift in SFT.
In our setting, the KL term serves the same purpose: 
it limits overfitting to the small analog circuit dataset 
and helps preserve the original general capabilities.
The effect of this combination is a controlled adaptation process achieving better knowledge gain.

\subsubsection{Mathematical Formulation and Validation}
To implement this approach, we adopt the following loss function:
\begin{equation}
\mathcal{L} = \mathcal{L}_{\text{CE}}(y_{\text{predict}}, y_{\text{label}}) + \lambda \cdot D_{\text{KL}}(p_{\text{predict}} \parallel p_{\text{ref}}), \label{eq:kl-regularized-sft}
\end{equation}
where $\mathcal{L}_{\text{CE}}$ is the standard cross-entropy loss in Eq. \eqref{eq:sft}, measuring the discrepancy between the prediction $y_{\text{predict}}$ and the label $y_{\text{label}}$;
$\lambda$ is the weighting coefficient balancing the two objectives;
and $D_{\text{KL}}(p_{\text{predict}} \parallel p_{\text{ref}})$ applies the KL divergence definition in Eq.~\eqref{eq:kl_divergence} to our SFT training scenario.

Specifically, we denote the trainable parameters of the current model and the reference model as 
$\theta$ and $\theta_0$, respectively. The KL divergence definition in Eq.~\eqref{eq:kl_divergence} is instantiated in our SFT training scenario as the following more detailed form:
\begin{equation}
D_{\text{KL}}(p_{\theta} \parallel p_{\theta_0}) = \sum_{v \in \mathcal{V}} p_{\theta}(v \mid x^{(k)}) \log \left[ \frac{p_{\theta}(v \mid x^{(k)})}{p_{\theta_0}(v \mid x^{(k)})} \right],
\end{equation}
where 
$v \in \mathcal{V}$ represents all possible tokens in the dictionary,
$p_{\theta}(v \mid x^{(k)})$ is the output probability assigned by the current model to token $v$ given input $x^{(k)}$, 
and $p_{\theta_0}(v \mid x^{(k)})$ is the output probability assigned by the reference model to the same token. 
This formulation compute the distribution difference at each token position between the current model and the reference model.

Empirical validation in Section \ref{s5} shows that this KL divergence regularization approach can achieve a 2.71 percentage-point improvement over traditional SFT for a 32B model, and for smaller 7B models, the improvement is even larger.

\subsection{Practical Implementation for Resource-constrained Scenarios}\label{s4d}
\subsubsection{The Challenge of Implementing SFT with KL Regularization}
In the EDA community, computational resources for LLM training are often limited. 
Therefore, beyond the algorithmic effectiveness, 
hardware feasibility of KL-regularized SFT should also be considered.
Unlike standard SFT, KL-regularized SFT requires maintaining two LLMs during training: 
a target model (abbreviated as Tar. in Fig. \ref{fig:data_flow}) to be optimized, 
and a frozen reference model (abbreviated as Ref. in Fig. \ref{fig:data_flow}) that provides the baseline distribution for the KL term. 
This means that an additional forward pass of the reference model must be executed,
and the outputs of both models must be retained simultaneously for KL divergence computation. 
This leads to the following challenges:

\textbf{(i) Without distributed training, GPU memory would overflow.} 
Consider our experimental setup: training a 32B target model at BF16 precision. 
The target model alone occupies 64GB for parameters. 
Keeping both target and reference models fully resident on each GPU doubles the requirement to 128GB. 
Adding gradients, optimizer states, activations, and communication buffers pushes the per-GPU memory demand beyond the 141GB limit of an advanced NVIDIA H200 GPU.

\textbf{(ii) Existing distributed SFT techniques are not readily applicable to dual-model scenarios.} 
Mainstream distributed SFT framework DeepSpeed ZeRO-3 \cite{deepspeed} registers hooks on each module to gather partitioned parameters before forward pass and release them after backward pass. 
This scheduling logic assumes a single model and a fixed execution order: forward, then backward. 
However, two models are used in this paper: 
a target model that is being trained, and a frozen reference model that only runs forward passes. 
The reference model triggers forward hooks but never enters backward pass, leaving gathered parameters in an unreleased state. 
Meanwhile, the interleaved execution of the two models sends conflicting requests to the same parameter state machine. 
ZeRO-3 cannot resolve these conflicts because its hook system was not designed for more than one model. 
Actually, naive attempts to run both models under ZeRO-3 resulted in frequent hangs or out-of-memory errors. 
It seems hard to directly use a ZeRO-3-like distributed training framework for this dual-model setup without modifications to its core C++ code.

\subsubsection{The Customized Implementation Scheme}

\begin{figure*}[!h]
    \centering
    \includegraphics[width=1\linewidth]{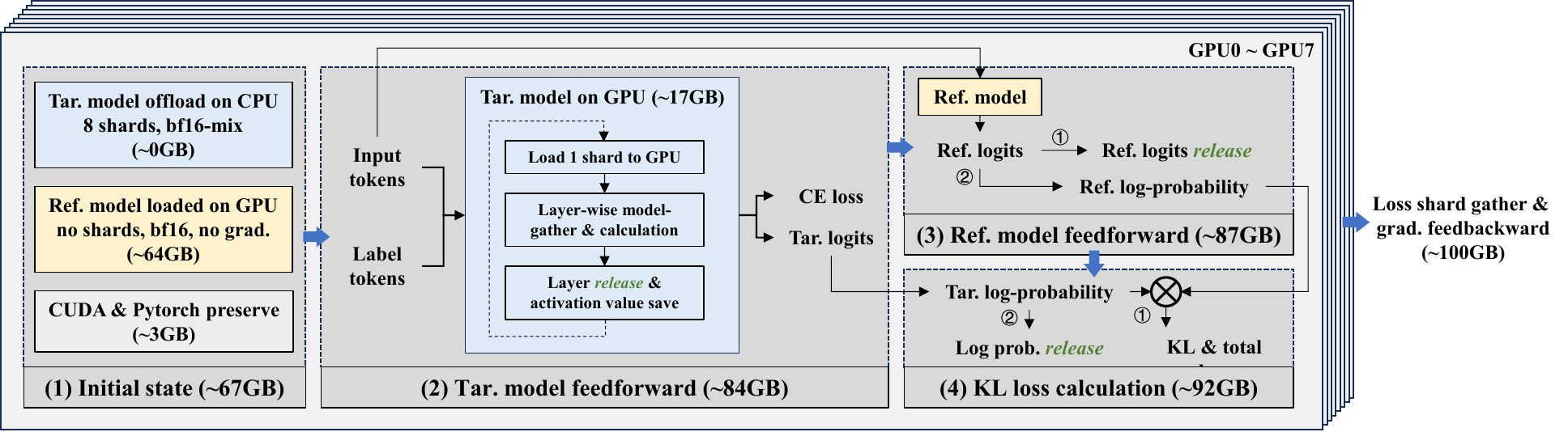}
\caption{The implementation of SFT with KL divergence regularization. This diagram specifically demonstrates the data flow during the forward propagation and gradient backpropagation process, along with the estimates of the memory usage per device for each step. The peak memory usage is tested to be 101.37GB per device.}
\label{fig:data_flow}
\end{figure*}

An asymmetric dual-model deployment scheme is proposed to address the above challenges, enabling SFT with KL regularization. 
Only the trainable target model is optimized using DeepSpeed ZeRO-3 sharding, while the frozen reference model is loaded in full on each GPU as a static local copy. 
Fig.~\ref{fig:data_flow} illustrates the overall approach. 
The following description uses a setup example: 
8 GPUs, a 32B model, BF16 precision, and a maximum sequence length of 8192 tokens.

\textbf{(i) Asymmetric dual-model deployment.}
The scheme applies DeepSpeed ZeRO-3 only to the target model. 
ZeRO-3 shards its parameters, gradients, and optimizer states across GPUs and offloads inactive parameters to CPU when possible, reducing per-GPU memory usage. 
In contrast, the reference model is frozen and only participates in forward inference for KL computation. 
It requires neither gradients nor optimizer states, so it is loaded entirely on each GPU without ZeRO-3 management. 
This asymmetric deployment introduces extra static memory cost but avoids the scheduling conflicts that arise when both models run under ZeRO-3. 
It enables data-parallel training (at least one sample per GPU) while keeping the implementation simple and stable.

\textbf{(ii) Target model forward pass.}
At the start of training, each GPU already holds the full reference model (64GB) with CUDA runtime overhead, resulting in an initial memory footprint of about 67GB. 
The target model then performs its forward pass under ZeRO-3. 
As shown in Fig.~\ref{fig:data_flow},
parameters are gathered layer by layer and released immediately after use, 
so the full set of training parameters never resides persistently on a single GPU.

\textbf{(iii) Cross-entropy loss computation.}
After the target model forward pass, 
the cross-entropy loss for SFT is computed. 
Intermediate activations that are not needed for subsequent KL computation are explicitly deleted to free memory. 
Only the target model output logits required for the KL computation are retained.

\textbf{(iv) Reference model forward pass.}
Once the target logits are obtained, 
the reference model performs its forward pass independently on each GPU. 
Because the reference model is frozen, no computation graph is built for backpropagation.
During this stage, per-GPU memory usage rises to approximately 87GB. 
Loading the reference model locally sacrifices some memory 
but simplifies the execution path and improves training stability.

\textbf{(v) KL divergence computation.}
With both target and reference model outputs available, 
the KL divergence is computed. 
At this point, logits from both models briefly coexist in memory, 
creating a temporary peak. 
For numerical stability, \texttt{log-softmax} is applied to each set of logits and the KL divergence is computed in log space. 
Immediately after computation, logits and temporary tensors are explicitly deleted. 
At this stage, per-GPU memory usage exceeds 90GB.

\textbf{(vi) Target model backward pass and parameter update.}
As defined in Eq.~\eqref{eq:kl-regularized-sft}, 
the total loss consists of the supervised cross-entropy term and the KL divergence term. 
Backpropagation and parameter updates are performed only on the target model. During this phase, gradient computation and ZeRO-3 parameter reconstruction cause memory usage to rise again. In practice, due to CUDA caching and framework-level preallocation, the per-GPU memory footprint stabilizes at 101.37GB, confirming that the scheme runs stably on a standard 8$\times$H200 server.

\subsubsection{Other details and applicability}
This scheme is realized through targeted customization at the \texttt{Python} training pipeline level.
First, at initialization, 
only the target model is handed over to the DeepSpeed engine; 
the reference model is loaded separately as a frozen static model on each GPU, bypassing ZeRO-3 scheduling. 
Second, the loss computation flow is rewritten: 
target model outputs are obtained first to compute the supervised loss, 
then reference model outputs are computed under \texttt{torch.no\_grad()}, 
and finally the KL divergence is calculated online and combined into the total loss. Third, to suppress transient memory spikes during long-sequence dual-model training, fine-grained tensor cleanup is inserted at stage boundaries, explicitly deleting intermediate variables that are no longer needed.

This scheme does not rely on modifying DeepSpeed's bottom C++/CUDA kernels. 
Instead, it builds upon existing training frameworks to enable stable and reproducible dual-model training with KL divergence regularization under resource-constrained conditions. 
For EDA researchers working on LLM applications, 
this scheme shares a feasible solution to train as large as 32B-parameter models with KL regularization on standard 8-GPU servers.

\section{EXPERIMENTS}\label{s5}
\subsection{Basic Environment} 
\subsubsection{Hardware Environment}
We utilize a server with 8 NVIDIA H200 SXM GPUs, each equipped with 141GB memory (700W), providing sufficient computational resources for medium-scale model training.
For circuit simulation, we use an Intel Xeon Platinum 8358 CPU. 

\subsubsection{Software Environment}
Our software infrastructure is built on CUDA 12.2, ensuring compatibility with the Transformers framework \cite{transformers} and DeepSpeed framework \cite{deepspeed}. All models are loaded using BF16 precision for memory efficiency, with flash-attention techniques \cite{flashattention,flashattention2} disabled to maintain numerical stability during KL-divergence calculations. 
For inference and evaluation, we leverage vLLM \cite{vllm} for high-throughput parallel processing.

\subsection{Training Settings} 
\subsubsection{Model Choice}
The training begins with Qwen 2.5-32B-Instruct, a medium-sized model with good baseline performance \cite{qwen2}. 
Ablation studies also use the equally sized reasoning model QwQ-32B \cite{qwq32b} and the Qwen 2.5-7B-Instruct model \cite{qwen2} for KL weight experiments.
\subsubsection{General-Domain Data Mixing}
During the SFT phase, we incorporate the OpenThoughts dataset \cite{openthoughts} with 20k samples alongside our analog-circuit domain data, maintaining a balanced scale between general and domain-specific samples, which is a common practice in domain-specific model training.
\subsubsection{Hyper-parameter Settings}
We use a cosine-annealing scheduler with a 10\% warm-up fraction, 
empirically determining that a maximum learning rate of $2\times10^{-6}$ works well for both CPT and SFT,
with modest variations showing negligible impact on performance.
Our training configuration allocates 1 sequence per GPU batch with 8-step gradient accumulation, resulting in a global batch size of 64 that ensures training stability. 
We set the maximum sequence length at 8,192 tokens and implement packing for shorter sequences to maximize computational efficiency.
The weighting coefficient $\lambda$ for KL divergence regularization is set to 0.1 according to the empirical training results on small-scale models.

\subsection{Evaluation Settings}
\subsubsection{AMSBench-TQA Benchmarking}
Trained models are evaluated on AMSBench-TQA, a third-party benchmark in \cite{amsbench}.
AMSBench-TQA is a human-crafted, text-only benchmark specifically designed to assess analog circuit knowledge through textual question-answering, which aligns with our research objective of evaluating pure textual knowledge understanding in analog circuits.
While AMSBench encompasses multiple subsets, the other subsets primarily focus on vision-language capabilities, which evaluate different competencies than our current research objectives.
AMSBench-TQA comprises a fixed set of multiple-choice problems, making automatic grading straightforward.
All trained models share a system prompt and a text generation temperature value of 0. 
To ensure fair evaluation, data leakage between the SFT dataset and the benchmark is examined. 
Using the embedding model from \cite{reimers2019sentence}, 
cosine similarities between the SFT dataset and the benchmark samples are calculated. 
Pairs with a similarity above 0.75 were manually checked to confirm whether 
any benchmark samples or their near variants were present in the SFT dataset. 
76 contaminated benchmark samples were identified and removed, yielding a deduplicated test set with 1181 samples in our experiment.

\subsubsection{Atelier Op-amp Design Task}
Additionally, the Atelier framework \cite{atelier} is employed to validate if the model can assist op-amp design.
Atelier is a multi-agent framework that combines LLM agents, a fixed workflow, and knowledge bases for circuit design. 
This framework requires the agent to perform topology selection, topology modification, and circuit analysis. 
Topology selection and modification refer to the agent choosing or adjusting the circuit structure based on design specs and simulation outcomes; 
circuit analysis requires the agent to refine the topology into a HSPICE netlist and specify the tunable parameters along with ranges. 
The design specification (spec) in the experiment is 
gain $>$ 85 dB, gain-bandwidth product (GBW) $>$ 5 MHz, phase margin (PM) $>$ 55 degrees, power $<$ 250 $\mu \text{W}$, and the comprehensive figure of merit (FoM) is expected to be large.
The maximum number of iterations is set to 4.

\subsection{Benchmarking Result Analysis}
Table \ref{table:performance_comparison} presents the performance of various baseline models and our model on this benchmark. 
The trained models are compared with commercial LLMs including Claude-Sonnet-4 \cite{anthropic2025claude4}, DeepSeek-R1 \cite{deepseek-r1}, DeepSeek-V3 \cite{deepseek-v3} and GPT-4o \cite{gpt4o}.
The open-source baseline model Qwen2.5-32B-Instruct \cite{qwen2}, 
the reasoning model QwQ-32B \cite{qwq32b} with a comparable scale,
and the recent Qwen3-32B model \cite{qwen3} are included for comparison.

\setlength{\tabcolsep}{3pt}
\renewcommand{\arraystretch}{1.3}
\begin{table}[!b]
    \centering
    \caption{Performance Comparison between Our Model and Baseline Models on AMSBench-TQA Benchmark}
    \label{table:performance_comparison}
    \small
    \begin{minipage}{0.92\linewidth}
    \centering
    \begin{tabular}{c|c|c|c|c}
        \hline
        \textbf{Starting point} & \textbf{Model scale} & \textbf{Method} & \textbf{Group} & {\textbf{Accuracy} (\%)} \\ \hline
        Claude-Sonnet-4      & Unknown    & —                 & Baseline      & 85.77 \\ \hline
        DeepSeek-R1          & 671B       & —                 & Baseline      & 84.93 \\ \hline
        Qwen2.5-32B-Instruct & 32B        & SFT+KL            & Experiment    & 84.59 \\ \hline
        DeepSeek-V3          & 671B       & —                 & Baseline      & 83.66 \\ \hline
        Qwen3-32B            & 32B        & —                 & Baseline      & 80.86 \\ \hline
        QwQ-32B              & 32B        & —                 & Baseline      & 80.78 \\ \hline
        GPT-4o               & Unknown    & —                 & Baseline      & 73.24 \\ \hline
        Qwen2.5-32B-Instruct & 32B        & —                 & Baseline      & 68.92 \\ \hline
    \end{tabular}
    \end{minipage}
\end{table}

\subsubsection{Comparison with Similar-Scale Baseline Models} \label{s4:comparison_with_similar_scale_baseline_models}

As shown in Table \ref{table:performance_comparison},
the model trained from Qwen2.5-32B-Instruct with KL-regularized SFT achieves an accuracy of 84.59\%, 
marking a 15.67 percentage-point improvement over the baseline instruct model Qwen2.5-32B-Instruct (68.92\%).
This improvement demonstrates the effectiveness of the proposed dataset and training method.

Besides, compared to Qwen2.5-32B-Instruct, the more advanced models from the same series,
QwQ-32B and Qwen3-32B, achieve accuracies of 80.78\% and 80.86\%, respectively,
and also outperform Qwen2.5-32B-Instruct (68.92\%).

QwQ-32B, which builds on the Qwen2.5-32B-Instruct baseline, improves its reasoning ability through large-scale RL \cite{qwq32b}. 
Although not specifically trained on analog circuit data, its strong reasoning capability allows it to derive correct answers through logical deduction, partly compensating for the lack of domain knowledge. 
For example, as shown in Fig. \ref{fig:example}, when analyzing noise in a classic CMOS op-amp, 
QwQ-32B applies basic formulas and correctly reasons that increasing $g_{\mathrm{m}1}$ reduces the first-stage noise contribution, while lowering the $g_{\mathrm{m}3}/g_{\mathrm{m}1}$ ratio increases first-stage gain and thus lowers input-referred noise.

Qwen3-32B, as a new-generation model, 
not only adopts a more advanced architecture,
but also adds trillions of tokens during the pre-training stage, with a particular increase in the STEM (science, technology, engineering, and mathematics) data \cite{qwen3}. 
This targeted training optimization enhances its overall ability,
explaining why Qwen3-32B achieves competitive performance despite lacking specialized training on analog circuit data.

\subsubsection{Comparison with Larger-Scale Commercial Models} \label{s5:comparison_with_larger_scale_commercial_models}
As shown in Table \ref{table:performance_comparison},
larger-scale baseline models, Claude-Sonnet-4, DeepSeek-R1 and DeepSeek-V3 achieve 85.77\%, 84.93\% and 83.66\% accuracy, respectively. 
The performance of our model is comparable to that of these larger-scale commercial models on this benchmark.

Especially, Claude-Sonnet-4, released later than DeepSeek-R1, is a newer model that generally outperforms DeepSeek-R1 in multiple evaluations \cite{anthropic2025claude4}.
Therefore, its top rank in Table \ref{table:performance_comparison} is expected.
On the other hand, GPT-4o's relatively ordinary performance (73.24\%) reflects its divergent training objectives, as GPT-4o is designed as a model for daily interaction \cite{gpt4o} rather than for professional technical assistance.

DeepSeek-V3 achieves 83.66\% accuracy. 
This performance is primarily attributed to two factors. 
First, DeepSeek-V3 adopts a Mixture-of-Experts architecture with 671B total parameters, 20.97 times larger than our 32B model \cite{deepseek-v3}. 
Second, the model is pre-trained on trillions of diverse and high-quality tokens, including substantial amounts of challenging math and code data \cite{deepseek-v3}. 
As the model used in our knowledge construction pipeline, 
DeepSeek-R1 further achieves 84.93\% accuracy, slightly higher than DeepSeek-V3 and our model. 
This improvement stems from the RL training applied to the 671B DeepSeek-V3, 
which enhances the model's reasoning capabilities through GRPO \cite{deepseek-r1}. 
Although Claude-Sonnet-4 achieves a slightly higher accuracy (85.77\%),
DeepSeek-R1 still delivers competitive performance on the benchmark.
Given our resource and budget limitations, 
DeepSeek-R1 is selected as the backbone LLM for data generation because of its reasonable balance between performance and cost.

\setlength{\tabcolsep}{3pt}
\renewcommand{\arraystretch}{1.3}
\begin{table}[!h]
    \centering
    \caption{Performance Comparison for Ablation Study I on AMSBench-TQA Benchmark}
    \label{table:ablation_i}
    \small
    \begin{minipage}{0.92\linewidth}
    \centering
    \begin{tabular}{c|c|c|c|c}
        \hline
        \textbf{Starting point} & \textbf{Model scale} & \textbf{Method} & \textbf{Group} & {\textbf{Accuracy} (\%)} \\ \hline
        Qwen2.5-32B-Instruct & 32B        & SFT+KL            & Experiment    & 84.59 \\ \hline
        Qwen2.5-32B-Instruct & 32B        & CPT+SFT+KL        & Ablation I    & 84.08 \\ \hline
        Qwen2.5-32B-Instruct & 32B        & CPT+SFT           & Ablation I    & 82.05 \\ \hline
        Qwen2.5-32B-Instruct & 32B        & SFT               & Ablation I    & 81.88 \\ \hline
        Qwen2.5-32B-Instruct & 32B        & CPT               & Ablation I    & 70.96 \\ \hline
        Qwen2.5-32B-Instruct & 32B        & —                 & Baseline      & 68.92 \\ \hline
    \end{tabular}
    \end{minipage}
\end{table}

\subsubsection{Ablation Study I: Analysis of CPT} \label{s5:ablation_i}
The CPT phase employs unsupervised domain texts for training, 
to enhance the model's understanding of analog circuit domain language patterns. 
As shown in Table \ref{table:ablation_i},
Qwen2.5-32B-Instruct (CPT) achieves 70.96\% accuracy, demonstrating that CPT alone yields a 2.04 percentage-point performance improvement for the original Qwen2.5-32B-Instruct. 
However, after subsequent SFT, the benefits of CPT become negligible. 
For example, Qwen2.5-32B-Instruct (CPT+SFT) achieves 82.05\% accuracy, 
only 0.17 percentage points higher than Qwen2.5-32B-Instruct (SFT).
Moreover, compared to Qwen2.5-32B-Instruct (SFT+KL), Qwen2.5-32B-Instruct (CPT+SFT+KL) underperforms by 0.51 percentage points. 

This phenomenon can be explained from two perspectives.
From a data scale perspective, as shown in Table \ref{table:dataset},
the CPT dataset is only 1/16 the size of the SFT dataset, resulting in a relatively limited impact on final model parameters. 
From a parameter update magnitude perspective, the supervised learning signal in the SFT phase is stronger  \cite{cpt,sft}, producing parameter updates far exceeding those from CPT, thereby overriding the subtle adjustments introduced by CPT in our scenario with imbalanced CPT and SFT data distribution.

\subsubsection{Ablation Study II. Analysis of KL Regularization} \label{s5:ablation_ii}

To demonstrate the performance improvement brought by the KL divergence regularization term in our experiments, repeated experiments were conducted using three different random seeds. 
In addition to the default seed 42, denoted as Seed A in this ablation study, 
we randomly selected seeds 123 and 9345, denoted as Seed B and Seed C, respectively, 
to perform comparative ablation experiments between SFT alone and SFT+KL. 
As shown in Table \ref{table:ablation_ii}, the KL divergence regularization term consistently improves performance over SFT across all three seeds, with an average improvement of 2.46 percentage points. 
Therefore, the KL divergence regularization term does improve the performance of SFT under this paper's analog circuit knowledge learning scenario.

\setlength{\tabcolsep}{3pt}
\renewcommand{\arraystretch}{1.3}
\begin{table}[!b]
    \centering
    \caption{Performance Comparison for Ablation Study II on AMSBench-TQA Benchmark}
    \label{table:ablation_ii}
    \small
    \begin{minipage}{0.92\linewidth}
    \centering
    \begin{tabular}{c|c|c|c|c}
        \hline
        \textbf{Starting point} & \textbf{Model scale} & \textbf{Method} & \textbf{Group} & {\textbf{Accuracy} (\%)} \\ \hline
        Qwen2.5-32B-Instruct & 32B        & SFT+KL            & Seed A       & 84.59 \\ \hline
        Qwen2.5-32B-Instruct & 32B        & SFT+KL            & Seed B       & 84.17 \\ \hline
        Qwen2.5-32B-Instruct & 32B        & SFT+KL            & Seed C       & 82.98 \\ \hline
        Qwen2.5-32B-Instruct & 32B        & SFT               & Seed A       & 81.88 \\ \hline
        Qwen2.5-32B-Instruct & 32B        & SFT               & Seed B       & 81.46 \\ \hline
        Qwen2.5-32B-Instruct & 32B        & SFT               & Seed C       & 81.03 \\ \hline
        Qwen2.5-32B-Instruct & 32B        & —                 & Baseline     & 68.92 \\ \hline

    \end{tabular}
    \end{minipage}
\end{table}

\begin{figure*}[!h]
    \centering
    \includegraphics[width=.9\linewidth]{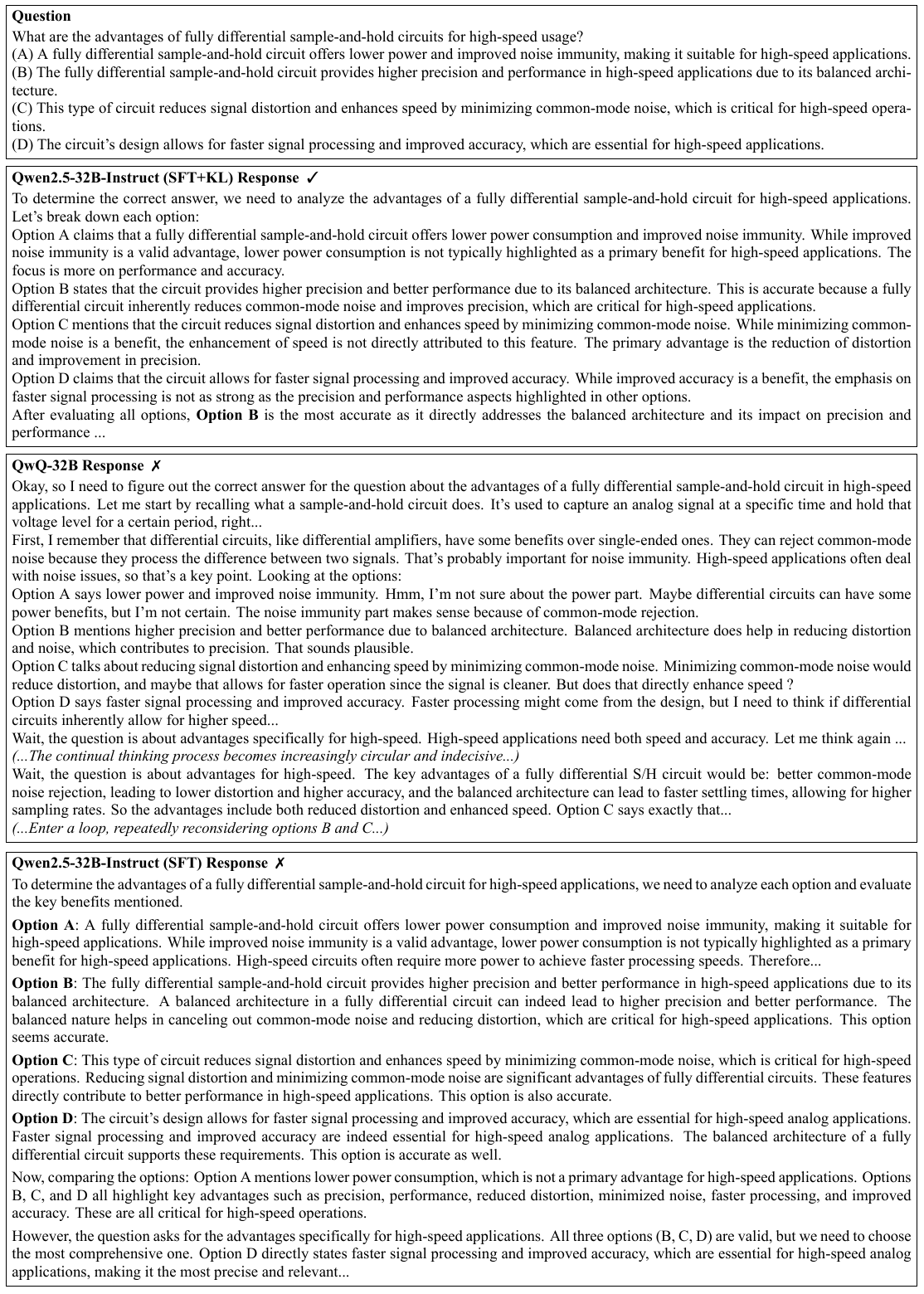}
    \caption{An example of models' response to a question in the AMSBench-TQA benchmark. 
    Note that for the same question, besides the model trained with SFT+KL and the original QwQ-32B, DeepSeek-V3, DeepSeek-R1 and Claude-Sonnet-4 also provide the correct answer. 
    Their logs are similar to those shown in this figure, and therefore, are not repeated here.}
    \label{fig:example}
\end{figure*}

Fig.~\ref{fig:example} provides a detailed example of this phenomenon. 
While Qwen2.5-32B-Instruct (SFT+KL) correctly analyzes CMOS op-amp noise and selects the correct option, 
the SFT-only model, Qwen2.5-32B-Instruct (SFT), makes a critical error: 
it claims the $g_{\mathrm{m}3}/g_{\mathrm{m}1}$ ratio only affects gain, not input-referred noise.
This is fundamentally incorrect. In multi-stage op-amps, the input-referred noise includes contributions from all stages, with later-stage noise referred to the input by dividing by the preceding gain. 
For a classical two-stage op-amp where the first-stage gain is approximately $g_{\mathrm{m}1}/g_{\mathrm{m}3}$, the total input-referred noise becomes:
$V_{n,in}^2 = V_{n1}^2 + V_{n3}^2 \cdot \left({g_{\mathrm{m}3}}/{g_{\mathrm{m}1}}\right)^2$,
where $V_{n1}^2$ and $V_{n3}^2$ are the noise power spectral densities of the first and second stages, $g_{\mathrm{m}1}$ and $g_{\mathrm{m}3}$ are the transconductances of the input and second-stage transistors.
This equation clearly shows that increasing $g_{\mathrm{m}1}/g_{\mathrm{m}3}$ reduces the second-stage noise contribution. 
The SFT-only model fails to recognize this noise referral mechanism and gets the relationship exactly backwards. 

This benchmark performance improvement brought by KL regularization can be explained from the following perspectives.
When fine-tuning for a specific domain via SFT, 
LLMs tend to overfit the small-scale target domain data, causing excessive parameter updates and leading to the forgetting of previously acquired capabilities such as language understanding, instruction following, and general reasoning \cite{forget,sft}. 
In the RL setting, numerous works \cite{dpo,ppo,deepseek-r1,rlrazor} have already demonstrated that a KL term can mitigate forgetting and steadily improve model performance.

In SFT, the cross-entropy loss provides the necessary gradient signal to inject domain-specific knowledge, while the KL regularization term acts as a mild constraint that prevents the model's parameter distribution from drifting too far from the reference model. 
This allows the model to absorb new knowledge while preserving its original capabilities as much as possible. Reference \cite{memoryretaining} experimentally demonstrates that adding a KL divergence term to the loss function can improve performance on the target domain without sacrificing other capabilities. 
Reference \cite{zhu2025anchored} further reports significant gains across tasks including mathematical reasoning, medical knowledge anchoring, and code generation by introducing a KL regularization term to curb excessive distribution shift during SFT.
These works align with the findings of this paper and jointly affirm the effectiveness of KL regularization for domain-specific SFT in analog circuit knowledge learning.

\subsubsection{Ablation Study III. KL Regularization Weight Settings} \label{s5:ablation_iii}

\setlength{\tabcolsep}{3pt}
\renewcommand{\arraystretch}{1}
\begin{table}[!b]
    \centering
    \caption{Performance Comparison for Ablation Study III on AMSBench-TQA Benchmark}
    \label{table:ablation_iii}
    \small
    \begin{minipage}{0.92\linewidth}
    \centering
    \begin{tabular}{c|c|c|c|c}
        \hline
        \textbf{Starting point} & \textbf{Model scale} & \textbf{Method} & \textbf{$\lambda$ value} & {\textbf{Accuracy} (\%)} \\ \hline
        Qwen2.5-7B-Instruct & 7B        & SFT+KL            & 0.00       & 55.63 \\ \hline
        Qwen2.5-7B-Instruct & 7B        & SFT+KL            & 0.01       & 60.80 \\ \hline
        Qwen2.5-7B-Instruct & 7B        & SFT+KL            & 0.05       & 77.30 \\ \hline
        Qwen2.5-7B-Instruct & 7B        & SFT+KL            & 0.08       & 79.42 \\ \hline
        Qwen2.5-7B-Instruct & 7B        & SFT+KL            & 0.09       & 79.42 \\ \hline
        Qwen2.5-7B-Instruct & 7B        & SFT+KL            & 0.10       & 79.85 \\ \hline
        Qwen2.5-7B-Instruct & 7B        & SFT+KL            & 0.11       & 78.41 \\ \hline
        Qwen2.5-7B-Instruct & 7B        & SFT+KL            & 0.12       & 78.49 \\ \hline
        Qwen2.5-7B-Instruct & 7B        & SFT+KL            & 0.20       & 77.98 \\ \hline
        Qwen2.5-7B-Instruct & 7B        & SFT+KL            & 0.30       & 77.82 \\ \hline
        Qwen2.5-7B-Instruct & 7B        & SFT+KL            & 0.40       & 77.22 \\ \hline
        Qwen2.5-7B-Instruct & 7B        & SFT+KL            & 0.50       & 76.55 \\ \hline
    \end{tabular}
    \end{minipage}
\end{table}

The weight of the KL divergence regularization term is commonly set around 0.1. 
To determine a suitable value of $\lambda$ for our experiments effectively,
the ablation study III on the 7B model is conducted. 
Fixing the random seed at 42, $\lambda$ values ranging from 0 to 0.5 are tested, with a particular focus around 0.1. 
As shown in Table \ref{table:ablation_iii},
when $\lambda$ increases from 0 to 0.1, model performance improves from 55.63\% to 79.85\%; when $\lambda$ further increases from 0.1 to 0.5, performance declines from 79.85\% to 76.55\%. 
This indicates that setting $\lambda$ to 0.1 achieves a balance between retaining general capabilities and acquiring new knowledge.

Moreover, for the 7B model, the introduction of KL divergence yields a performance improvement of 24.22 percentage points compared with SFT alone, which is larger than the 2.71 percentage-point gain for the 32B model.
This is because smaller LLMs have limited parameter capacity and are more susceptible to losing general capabilities during fine-tuning, as parameter updates tend to overwrite previously acquired knowledge. 
As noted in references \cite{kalajdzievski2024scaling,barron2025too}, small-scale models possess a very narrow tolerance margin for knowledge retention; unconstrained SFT can cause damage to their general capabilities.
In contrast, larger LLMs such as the 32B variant possess adequate parameter redundancy, 
and can experience less forgetting and overfitting during SFT \cite{barron2025too,kalajdzievski2024scaling}.
Consequently, larger LLMs can tolerate parameter updates induced by SFT to a certain extent without excessively compromising their original capabilities, and the benefit of KL regularization for larger LLMs is correspondingly smaller.

\subsubsection{Ablation Study IV: Analysis of Reasoning Models} \label{s4:ablation_iv}
QwQ-32B is a reasoning model optimized through large-scale RL \cite{qwq32b}, 
with its parameter distribution finely tuned to handle complex general reasoning tasks. 
The ablation study IV shown in Table~\ref{table:ablation_iv} reveals a consistent phenomenon: 
any domain-specific fine-tuning applied to this model leads to performance degradation in our scenario. 
Across experiments,
the maximum accuracy drop reaches 7.88 percentage points. 
SFT leads to substantial performance collapse regardless of whether KL divergence regularization is employed and whether CPT is used before SFT.

\setlength{\tabcolsep}{3pt}
\renewcommand{\arraystretch}{1}
\begin{table}[!h]
    \centering
    \caption{Performance Comparison for Ablation Study IV on AMSBench-TQA Benchmark}
    \label{table:ablation_iv}
    \small
    \begin{minipage}{0.92\linewidth}
    \centering
    \begin{tabular}{c|c|c|c|c}
        \hline
        \textbf{Starting point} & \textbf{Model scale} & \textbf{Method} & \textbf{Group} & {\textbf{Accuracy} (\%)} \\ \hline
        QwQ-32B & 32B        & —                 & Baseline       & 80.78 \\ \hline
        QwQ-32B & 32B        & SFT+KL            & Ablation IV    & 74.51 \\ \hline
        QwQ-32B & 32B        & SFT               & Ablation IV    & 74.34 \\ \hline
        QwQ-32B & 32B        & CPT+SFT+KL        & Ablation IV    & 73.07 \\ \hline
        QwQ-32B & 32B        & CPT+SFT           & Ablation IV    & 72.90 \\ \hline
    \end{tabular}
    \end{minipage}
\end{table}

These results indicate that reasoning models exhibit sensitivity to subsequent parameter updates, 
and injecting additional domain knowledge into such models is difficult. 
This phenomenon can be attributed to the fragility of parameter distributions induced by RL optimization.
Reasoning models like QwQ-32B undergo extensive online sampling and reward feedback during large-scale RL training~\cite{qwq32b,deepseek-r1,dapo}. 
Such high-intensity RL training of LLMs can cause the mapping from reward to optimal policy to be discontinuous rather than smooth, driving the model to converge to steep local parameter regions~\cite{policycliff}, where further parameter perturbations can trigger discontinuous shifts in output behavior, rendering these models resistant to further domain knowledge injection~\cite{kirk2023understanding,rlvssft,diao2026entropy}.

\subsubsection{Ablation Study V: Analysis of General Data and Domain Data} 
\label{sec:ablation_v}
To demonstrate the effectiveness of the data used in this paper, 
ablation study V was conducted. 
The results are presented in Table~\ref{table:ablation_v}. 

Training solely with analog circuit domain data yields a substantial performance improvement, achieving an accuracy of 82.81\%, which represents a gain of 13.89 percentage points over the baseline. 
This result indicates that the proposed SFT dataset is independently effective for injecting analog circuit domain knowledge, and does not rely on the capabilities of the general dataset.

General data was included not to inject domain knowledge, but to help the model retain its general reasoning and instruction-following abilities during domain training~\cite{ding2025improved}.
Including a mixture of general and domain data is standard practice in domain-specific LLM development~\cite{ling2025diversity,ming2025ideal}. 
The OpenThoughts dataset~\cite{openthoughts} contains general reasoning examples across mathematics, coding, and science, and is known to improve reasoning performance.
Therefore, Qwen2.5-32B-Instruct, a non-reasoning model, improved to 79.93\% after training on OpenThoughts, 
without injecting domain knowledge.
However, this gain remains lower than that achieved with domain data alone, 
confirming that the domain dataset provides information not captured by general reasoning data.

Mixed training with both general and domain data achieves the highest accuracy (84.59\%), 
which is 1.78 percentage points above the domain-only result.  
This gain reflects the different roles of the two data sources: 
domain data provides task-specific knowledge, 
while general data helps maintain the model's general capabilities and reduces overfitting to the domain distribution. 
References~\cite{ling2025diversity,ming2025ideal} indicate that diversity in training data contributes to improving overall LLM capabilities, particularly in scenarios with imbalanced domain data distributions, 
where diversity itself serves as an effective signal to prevent overfitting. 
This is also consistent with our results: 
combining general and domain data preserves general ability while incorporating domain knowledge, 
leading to the best overall performance.

In summary, this ablation study demonstrates that the domain dataset constructed in this work plays an independent and more important role in knowledge injection.

\setlength{\tabcolsep}{3pt}
\renewcommand{\arraystretch}{1}
\begin{table}[!t]
    \centering
    \caption{Performance Comparison for Ablation Study V on AMSBench-TQA Benchmark}
    \label{table:ablation_v}
    \small
    \begin{minipage}{1\linewidth}
    \centering
    \begin{tabular}{c|c|c|c|c}
        \hline
        \textbf{Starting point} & \textbf{Model scale} & \textbf{Method} & \textbf{Group} & {\textbf{Accuracy} (\%)} \\ \hline
        Qwen2.5-32B-Instruct & 32B        & SFT+KL               & Mixed Data          & 84.59 \\ \hline
        Qwen2.5-32B-Instruct & 32B        & SFT+KL               & Only Domain Data    & 82.81 \\ \hline
        Qwen2.5-32B-Instruct & 32B        & SFT+KL               & Only General Data   & 79.93 \\ \hline
        Qwen2.5-32B-Instruct & 32B        & —                    & Baseline            & 68.92 \\ \hline
    \end{tabular}
    \end{minipage}
\end{table}

\subsubsection{Summary}
Based on the aforementioned analysis, our experiments lead to the following findings:
(i) SFT with KL divergence regularization outperforms traditional SFT in adapting to the domain of analog circuit knowledge. 
(ii) In the context of analog circuits, the imbalanced data distribution between the CPT and SFT stages makes direct SFT much more effective than CPT.
(iii) Model selection is pivotal for achieving training success, with general instruct models proving to be more suitable as starting points for domain adaptation than highly specialized reasoning models.

\subsection{Agentic Op-amp Design Task Study} 
\label{sec:atelier_eval}

To further evaluate whether the trained models can serve as design assistants, 
we integrate the LLMs trained from Qwen2.5-32B-Instruct with different configurations into the Atelier op-amp design framework \cite{atelier}. 

\begin{figure*}[!t]
    \centering
    \includegraphics[width=1\linewidth]{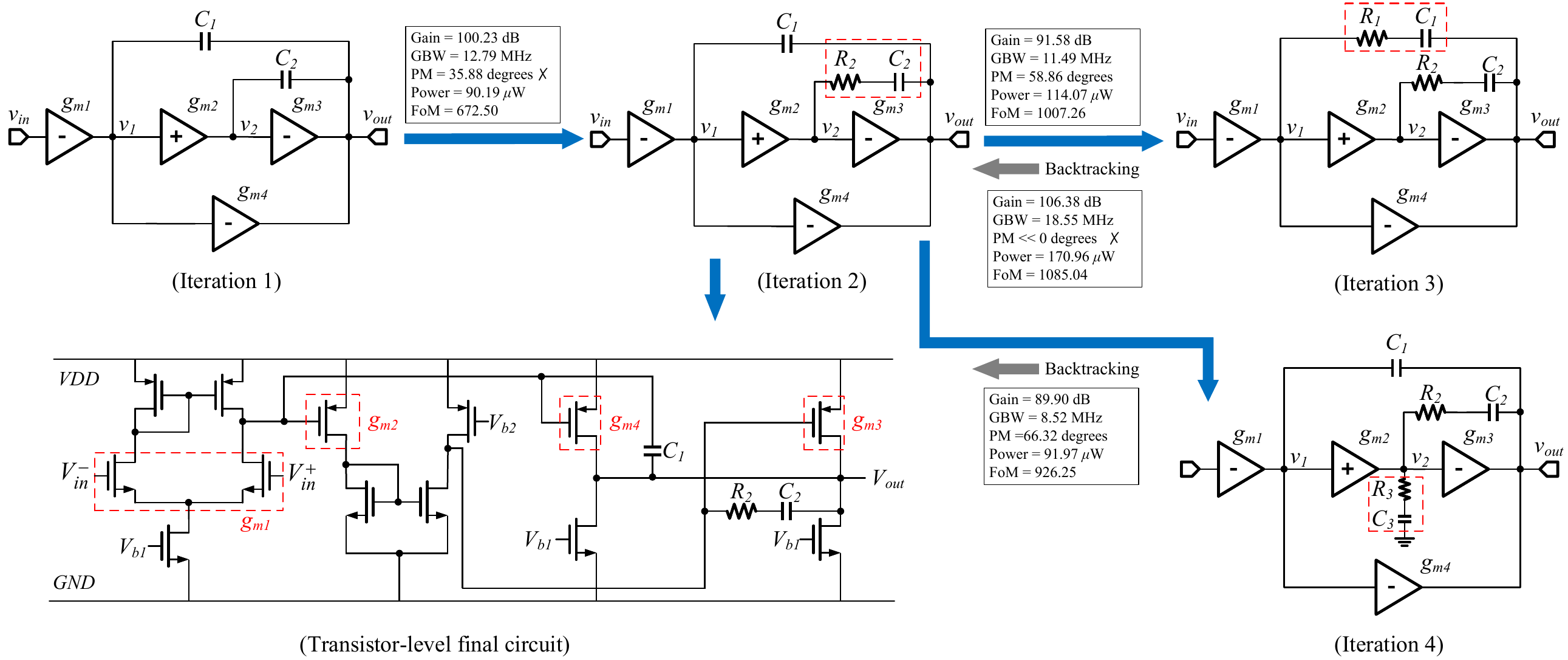}
    \caption{The op-amp design trajectory of Qwen2.5-32B-Instruct (SFT+KL) under the Atelier framework. For each iteration, the circuit designed by the agent and its corresponding performance metrics are presented.}
    \label{fig:trajectory}
\end{figure*}

\begin{figure*}[!t]
    \centering
    \includegraphics[width=1\linewidth]{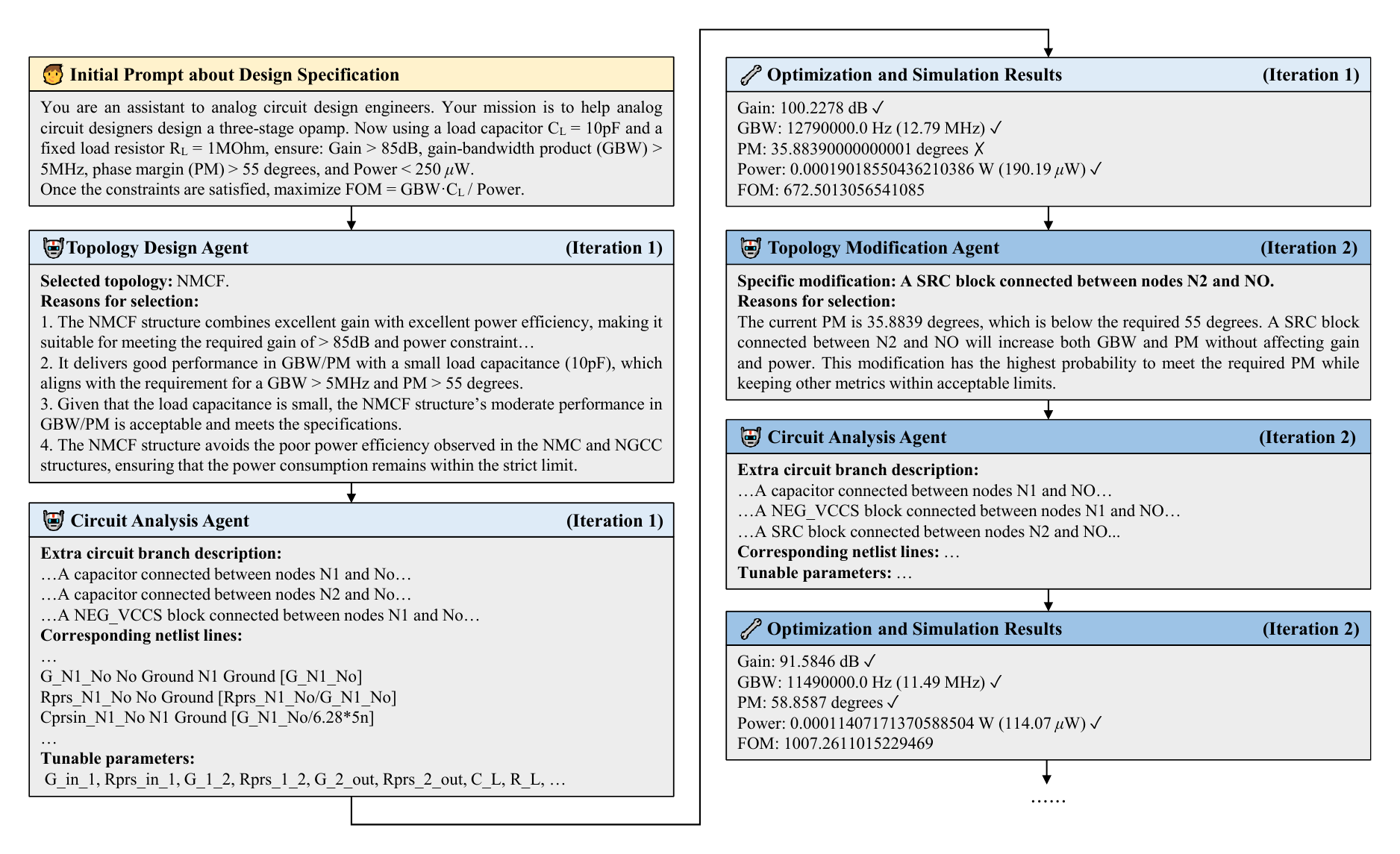}
    \caption{The chatlog of Qwen2.5-32B-Instruct (SFT+KL) within the Atelier op-amp design framework. The model assumes the roles of two agents: a topology design/modification agent and a circuit analysis agent. The detailed dialogue of the first two iterations is shown here. Subsequent content is omitted, but the overall workflow can still be seen in Fig. \ref{fig:trajectory}.
    Besides, due to space limitations, some details of the content have been omitted using ellipses.}
    \label{fig:chatlog}
\end{figure*}

Experimental results are presented in Table~\ref{table:atelier_op-amp_results}. 
The model trained with a mixture of domain and general data achieves a design success rate of 7/10 and an average figure of merit (FoM) of 533.48, representing a 42.5\% improvement over the baseline. 
Taking the design trajectory illustrated in Fig.~\ref{fig:trajectory} and Fig.~\ref{fig:chatlog} as an example: in the first iteration, the agent selects the NMCF topology, and the circuit analysis module converts it into a netlist followed by parameter optimization. 
At this stage, the PM is only 35.88°, indicating insufficient stability. 
In the second iteration, the agent introduces a nulling resistor to improve PM, and the modified design satisfies all requirements with an FoM of 1107.26. 
In the subsequent two iterations, the agent continues to explore alternative topological adjustments. 
These observations indicate that the trained model Qwen2.5-32B-Instruct (SFT+KL) can understand specs, leverage domain knowledge to select appropriate topologies, and make targeted adjustments based on feedback, demonstrating capabilities of a design assistance agent.

\setlength{\tabcolsep}{4pt}
\renewcommand{\arraystretch}{1}
\begin{table}[!h]
    \centering
    \caption{Atelier Op-amp design Task Results}
    \label{table:atelier_op-amp_results}
    \small
    \begin{minipage}{0.92\linewidth}
    \centering
    \begin{tabular}{l|c|c|c|c|c|c}
        \hline
        \textbf{Experimental group} & \textbf{Success rate} & \textbf{Gain (dB)} & \textbf{GBW (MHz)} & \textbf{PM ($^\circ$)} & \textbf{Power ($\mu$W)} & \textbf{FoM} \\
        \hline
        Full pipeline               & 7/10  & 75.94 & 7.47 & 40.39 & 109.19 & 533.48 \\
        General data only              & 7/10  & 66.67 & 5.45 & 40.19 & 91.59  & 451.96 \\
        Domain data only            & 6/10  & 57.83 & 5.41 & 35.39 & 91.41  & 378.06 \\
        baseline                    & 7/10  & 65.40 & 5.54 & 42.18 & 102.49 & 374.42 \\
        \hline
    \end{tabular}
    \end{minipage}
\end{table}

The model trained only on domain data achieves a success rate of 6/10 and an average FoM of 378.06, which is marginally improved from the baseline while the success rate slightly declines. 
This trend aligns with the fact that 
fine-tuning solely on domain data tends to cause degradation of general instruction-following capabilities \cite{ming2025ideal,ling2025diversity}. 
In our experiment, failure cases in this group usually involve netlist syntax errors or illegal output formats, suggesting that weaknesses in general abilities most directly affect the agent's performance when executing real tasks.

The model trained exclusively on the OpenThoughts general dataset achieves a success rate of 7/10 and an average FoM of 451.96, a 20.7\% improvement over the baseline, but still lower than the domain-general mixed group. 
OpenThoughts \cite{openthoughts} encompasses reasoning tasks across mathematics, programming, and science; 
its role is not to inject analog circuit expertise but rather to help the model preserve and improve general instruction-following and reasoning capabilities \cite{ding2025improved}. 
Consequently, despite lacking domain knowledge, 
this model can still successfully conduct design tasks by relying on general reasoning and by producing correctly formatted outputs, avoiding frequent instruction-following errors.

The domain-general mixed group attains the highest FoM (533.48) while maintaining the success rate, demonstrating the effective integration of domain knowledge and general capabilities. 
This result is consistent with prior work \cite{ling2025diversity,ming2025ideal} indicating that interleaving general data with domain data helps balance knowledge acquisition and capability retention, preventing excessive deviation from the original model distribution and yielding better performance in tasks.

\section{CONCLUSION}\label{s6}
This paper constructs a textual dataset for LLMs to learn analog circuit knowledge and customizes training approaches.
A dataset consisting of 7.26M tokens for CPT and 112.65M tokens for SFT is constructed, where structured QTSA quadruples are produced for the SFT stage to capture domain knowledge.
Several findings that may help researchers facing similar challenges are: 
instruct models serve as suitable initialization points, 
SFT-centric approaches outperform other pipelines in our setting, 
and KL divergence regularization can achieve a 2.71 percentage-point improvement over SFT alone.
Experiments demonstrate that a 32B instruct model trained on our dataset achieves 84.59\% accuracy on AMSBench-TQA, showing a 15.67 percentage-point improvement over the initial model, and demonstrates capability in the downstream op-amp design task.

\section{Acknowledgments}
This work is partially supported by the National Natural Science Foundation of China under research project 92373207, and by the Science and Technology Commission of Shanghai Municipality under grant 25JD1400400.
\bibliographystyle{ACM-Reference-Format}
\bibliography{ref}


\end{document}